\documentclass[aps,amsmath, prb,twocolumn,superscriptaddress]{revtex4-2}
\usepackage{lineno}
\usepackage{graphicx}
\usepackage{dcolumn}
\usepackage{bm}
\usepackage{lipsum}
\usepackage{color}
\usepackage{amsfonts,eucal,mathrsfs,amssymb,gensymb}

\usepackage[table]{xcolor}
\graphicspath{{./figs/}}
\usepackage{multirow}
\usepackage[breaklinks=true,unicode=true,urlcolor = blue,colorlinks = true,citecolor = blue,linkcolor = blue]{hyperref}

\usepackage[normalem]{ulem}

\renewcommand{\vec}[1]{\bm{#1}}
\newcommand{\dd}{\mathrm{d}}

\begin{document}

\preprint{}

\title{Nonlinear Magnon Magnetic Moment Transport in Triangular-Lattice f-Wave Antialtermagnets}


\author{Volodymyr P. Kravchuk}
\email{v.kravchuk@ifw-dresden.de}
\affiliation{Institute for Theoretical Solid State Physics, Leibniz Institute for Solid State and Materials Research Dresden, D-01069 Dresden, Germany}
\affiliation{Institute of Theoretical Physics and W{\"u}rzburg-Dresden  Cluster of Excellence {\it ctd.qmat}, Technische Universit{\"a}t Dresden, 01062 Dresden, Germany} 
\affiliation{Bogolyubov Institute for Theoretical Physics of the National Academy of Sciences of Ukraine, 03143 Kyiv, Ukraine}

\author{Kostiantyn V. Yershov}\affiliation{Institute for Theoretical Solid State Physics, Leibniz Institute for Solid State and Materials Research Dresden, D-01069 Dresden, Germany}
\affiliation{Bogolyubov Institute for Theoretical Physics of the National Academy of Sciences of Ukraine, 03143 Kyiv, Ukraine}

\author{Bastián Pradenas}\affiliation{Institute for Theoretical Solid State Physics, Leibniz Institute for Solid State and Materials Research Dresden, D-01069 Dresden, Germany}

\author{Robin R. Neumann}
\affiliation{Institute of Solid State Theory, University of Münster, D-48149 Münster, Germany}
\affiliation{Institute of Physics and Halle–Berlin–Regensburg Cluster of Excellence (CCE), Martin Luther University Halle–Wittenberg, D-06099 Halle (Saale), Germany}

\author{Rodrigo Jaeschke-Ubiergo}
\affiliation{Institute of Physics, Johannes Gutenberg University Mainz, D-55128 Mainz, Germany}

\author{Ricardo Zarzuela}
\affiliation{Institute of Physics, Johannes Gutenberg University Mainz, D-55128 Mainz, Germany}

\author{Jairo Sinova}
\affiliation{Institute of Physics, Johannes Gutenberg University Mainz, D-55128 Mainz, Germany}
\affiliation{Department of Physics, Texas A\&M University, College Station, Texas 77843-4242, USA}

\author{Jeroen van den Brink}
\affiliation{Institute for Theoretical Solid State Physics, Leibniz Institute for Solid State and Materials Research Dresden, D-01069 Dresden, Germany}
\affiliation{Institute of Theoretical Physics and W{\"u}rzburg-Dresden  Cluster of Excellence {\it ctd.qmat}, Technische Universit{\"a}t Dresden, 01062 Dresden, Germany} 

\author{Alexander Mook}
\affiliation{Institute of Solid State Theory, University of Münster, D-48149 Münster, Germany}

\begin{abstract}
We study the spin excitations in the frustrated coplanar 120$\degree$ ground state of the triangular-lattice Heisenberg antiferromagnet and demonstrate that they carry a magnetic moment perpendicular to the plane in which the spins order, despite the ground-state sublattice moments having no out-of-plane component. 
The symmetry of the momentum dependence of the magnetic moment and energy of the magnons renders the system an odd-parity $f$-wave magnet. 
Extending this model to a stack of antiferromagnetically coupled triangular layers provides a realization of magnons in a three-dimensional $f$-wave antialtermagnet. We show that nonlinear thermal transport effects of magnons, such as Edelstein and spin-splitter effects, provide clear experimental signatures of magnons in $f$-wave antialtermagnets.
\end{abstract}

\maketitle

\section{Introduction}
\begin{figure}
    \centering
    \includegraphics[width=\columnwidth]{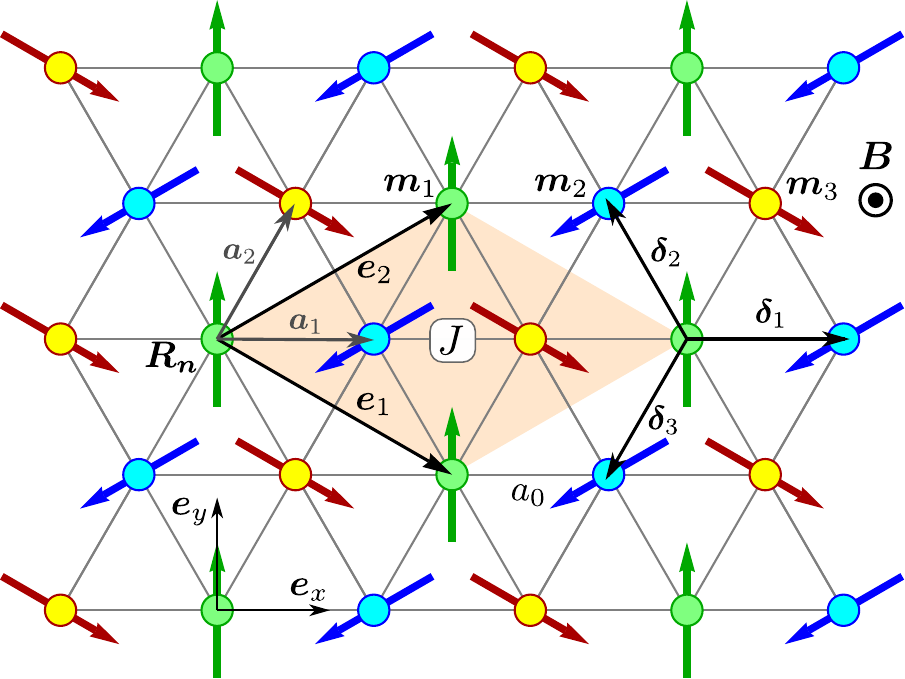}
    \caption{Triangular lattice of frustrated spins with $J>0$. Three uniformly magnetized sublattices $\vec{m}_1$, $\vec{m}_2$, and $\vec{m}_3$ are shown by different colors. The magnetic unit cell is determined by the vectors $\vec{e}_1$ and $\vec{e}_2$ and is shown by the orange shading, while $\vec{a}_1$ and $\vec{a}_2$ represent the primitive basis vectors of the triangular lattice. The magnetic field $\vec{B}$ is applied perpendicularly to the lattice.}
    \label{fig:lattice}
\end{figure}

 A combination of spin-frustration and low dimensionality gives rise to a variety of new physical effects~\cite{Diep13,Lacroix11}. The triangular-lattice Heisenberg antiferromagnet is a canonical model of geometric frustration: for nearest-neighbor antiferromagnetic exchange, spins cannot satisfy all bonds simultaneously. In the limit of zero temperature and for the vanishing magnetic field, the ground state is a coplanar three-sublattice ``120-degree'' state~\cite{Kawamura84,Miyashita86,Chubukov91,Capriotti99,Seabra11}, see Fig.~\ref{fig:lattice}.  Remarkably, this ground state is not destroyed by the quantum fluctuations even for spin-$1/2$ systems~\cite{Jolicoeur89,Chubukov94,Capriotti99,White07a}.  The spectra of spin waves excited on the top of the 120$\degree$-state were extensively studied analytically~\cite{Jolicoeur89,Chernyshev09,Mourigal13,Maksimov16,Syromyatnikov22}, numerically~\cite{Zheng06a,Zheng06}, and experimentally~\cite{Gaulin87,Bordelon20,Ma16b}. It is also worth noting the works in which a continuous field theory was developed that describes low-energy excitations (linear as well as nonlinear) of the 120$\degree$-state~\cite{Dombre89,Dasgupta20,Pradenas24,Pradenas25,Zarzuela25a}.
 

In many simple cases, e.g., in easy-axial antiferromagnets, the direction of spin precession in the excited spin wave is a well-defined property of a given magnon branch, and the terminology of chiral magnons is used in this context~\cite{Kravchuk25a,Smejkal23,Wang24}. In the general case, however, the spins in different sublattices can precess in opposite directions, and/or demonstrate linear polarization in some of the sublattices~\cite{Kravchuk25a}. In these cases, the concept of magnon chirality is ill-defined, so we use the magnon magnetic moment
 \begin{equation}\label{eq:mu}
     \vec{\mu}_{\nu,\vec{k}}=-\left.\partial_{\vec{B}}\varepsilon_{\nu,\vec{k}}\right|_{\vec{B}=\vec{0}},
 \end{equation}
which is well-defined in the general case. Here $\varepsilon_{\nu,\vec{k}}$ is the energy of the magnon of $\nu$-th branch with the wave vector $\vec{k}$, and $\vec{B}$ is the applied magnetic field. With three sublattices, we have three magnon branches, which we number as $\nu=-1,0,1$ for convenience. The relation~\eqref{eq:mu} can be obtained by considering the total magnetic moment $\vec{M}$ of the Bose-gas of magnons as a thermodynamic variable $\vec{M}=-(\partial F/\partial\vec{B})_{T,V}$ with $F$ being the Helmholtz free energy~\cite{Ashcroft76,Aharoni96,Neumann20,Yershov24b}. 
For the triangular-lattice Heisenberg antiferromagnet we find that $\vec{\mu}_{\nu,\vec{k}}=\mu^z_{\nu,\vec{k}}\vec{n}$, where $\vec{n}$ is a unit vector perpendicular to the plane in which spins order. In what follows, without loss of generality, we assume that $\vec{n}=\hat{\vec{z}}$, where ort $\hat{\vec{z}}$ is orthogonal to the $xy$-plane in which the lattice lies, see Fig.~\ref{fig:lattice}. This is rather counterintuitive, since none of the sublattices are magnetized in the $z$-direction. Therefore, the magnetic moment~\eqref{eq:mu} cannot be interpreted as a fraction of the ground state magnetization of a sublattice taken away by a magnon. The latter interpretation is commonly used for the explanation of Bloch's $T^{3/2}$-law for ferromagnets~\cite{Ashcroft76}. Nevertheless, the collinear spin polarization in the momentum space, which is directed perpendicularly to coplanar spins in the position space, is typical for the coplanar odd-parity wave magnets~\cite{Hellenes23,Mitscherling26} dubbed antialtermagnets~\cite{Jungwirth25}. The latter are the odd-parity counterparts to the widely discussed altermagnets~\cite{Smejkal22a,Smejkal22b}.

We find that in the nonrelativistic limit, the spectrum of magnons excited over the 120$\degree$-state possesses symmetries
\begin{equation}\label{eq:sym}
\varepsilon_{\nu,\vec{k}}=\varepsilon_{-\nu,-\vec{k}},\qquad \mu^z_{\nu,\vec{k}}=-\mu^z_{-\nu,-\vec{k}},
\end{equation}
which is typical for odd-parity magnets~\cite{Hellenes23,Jungwirth25,Neumann26}. While the general properties of the odd-parity-wave magnons for noncoplanar magnets were recently analyzed in Ref.~\onlinecite{Neumann26}, here we consider the 120$\degree$ antiferromagnetic state as a case study of a \emph{coplanar} odd-parity-wave magnet. 
In particular, we demonstrate that the magnon spectrum possesses f-wave symmetry with respect to the magnetic moment defined in Eq.~\eqref{eq:mu}, and elucidate the Zeeman splitting of the oppositely magnetized magnon modes. We confirm our analytical findings by means of numerical spin-lattice simulations: by reconstructing the dynamics of individual spins, we investigate the `mechanics' of the magnon modes in the low-energy regime and determine the influence of the magnetic field upon them.

Next, we consider a 3D extension of our model as an infinite stack of monolayers. The multilayer model possesses the $T\vec{t}$-symmetry~\footnote{Here $T$ and $\vec{t}$ denote time reversal and translation, respectively.} with the translation along $z$-axis, and therefore satisfies the defining symmetry criterion of an antialtermagnet~\cite{Hellenes23,Jungwirth25}. For this system, we predict two nonrelativistic thermally induced nonlinear transport effects of magnons: spin-splitter effect~\cite{Yershov24b,Weissenhofer24,Cui23a,Yang26b}, and Edelstein effect~\cite{Zhang20b,Li20f}, which are quadratic and cubic in the temperature gradient, respectively. The first of these effects is the generation of a flow of magnetic moment carried by magnons, which we will also call spin current, in response to an applied temperature gradient. This effect was theoretically predicted~\cite{Yershov24b,Weissenhofer24,Cui23a} and recently observed experimentally~\cite{Yang26b} for $d$-wave altermagnets. The direction of the generated spin current depends on the direction of the applied temperature gradient relative to the crystallographic directions~\cite{Yershov24b,Weissenhofer24,Cui23a,Cui26}: it can be transverse or longitudinal, mimicking the spin Nernst or spin Seebeck effect geometries, respectively. 

\section{Spin waves in a single layer}
Although the spectrum of the linear spin waves excited on top of the 120$\degree$ antiferromagnetic state was previously obtained~\cite{Jolicoeur89,Chernyshev09,Mourigal13,Maksimov16,Syromyatnikov22}, including the case with the applied magnetic field~\cite{Maksimov16}, in this section, we focus on the magnetic moment~\eqref{eq:mu} carried by the magnons and analyze in detail the structure of the magnon modes in terms of the single spin dynamics. As a model, we consider a two-dimensional triangular lattice of discrete magnetic moments described by the classical Heisenberg Hamiltonian with a magnetic field
\begin{equation}\label{eq:H}
    \mathcal{H}=\frac{J}{2}\sum\limits_{\langle\vec{r}_{\vec{n}},\vec{r}'_{\vec{n}}\rangle}\vec{m}(\vec{r}_{\vec{n}})\cdot\vec{m}(\vec{r}'_{\vec{n}})-\mu_sB\sum\limits_{\vec{r}_{\vec{n}}}m_z(\vec{r}_{\vec{n}}).
\end{equation}
Here, $\vec{m}(\vec{r}_{\vec{n}})=\vec{\mu}(\vec{r}_n)/\mu_s$ with $\mu_s=|\vec{\mu}(\vec{r}_{\vec{n}})|$ is a unit vector in the direction of the magnetic moment located in the node $\vec{r}_{\vec{n}}=n_1\vec{a}_1+n_2\vec{a}_2$. Furthermore, $n_i\in\mathbb{Z}$ and $\vec{a}_1=a_0\hat{\vec{x}}$, $\vec{a}_2=a_0[\hat{\vec{x}}+\sqrt{3}\hat{\vec{y}}]/2$ are the primitive basis vectors which correspond to the structural (crystal) unit cell, see Fig.~\ref{fig:lattice}. Finally, $a_0$ is the spacing between the nearest neighbors, $\vec{r}'_{\vec{n}}$ labels the six nearest neighbors of $\vec{r}_{\vec{n}}$, and $J>0$ is the antiferromagnetic exchange. Without the external magnetic field, the ground state of $\mathcal{H}$ is the coplanar 120$\degree$ state with the arbitrarily oriented plane containing the magnetic moments. Without loss of generality, one can assume that the spin plane is the $xy$-plane. Application of the perpendicular magnetic field $\vec{B}=B\hat{\vec{z}}$ cants all spins towards the field, forming the ``umbrella'' state in which all spins acquire a component $m_z=B/B_0$ along the field, where $B_0=9J/\mu_s$ is the spin-flip field. Assuming that the dynamics of each magnetic moment is governed by the Landau-Lifshitz equation 
\begin{equation}\label{eq:LL}
    \dot{\vec{m}}(\vec{r}_{\vec{n}})=\frac{\gamma}{\mu_s}\left[\vec{m}(\vec{r}_{\vec{n}})\times\frac{\partial\mathcal{H}}{\vec{m}(\vec{r}_{\vec{n}})}\right]
\end{equation}
with $\gamma=|g|\mu_{\textsc{b}}/\hbar>0$ being the gyromagnetic ratio, we linearize the set of equations arising from the vector components of Eq.~\eqref{eq:LL} in the vicinity of the ``umbrella'' state and, using the Holstein-Primakoff formalism, we solve the corresponding eigenvalue problem on the discrete lattice; for details, see Appendix~\ref{app:spectra}. The obtained spectra of the magnon eigenenergies $\varepsilon_{\nu,\vec{k}}$ are as follows:
\begin{align}\label{eq:E_k}
    \nonumber&\varepsilon_{\pm1,\vec{k}}\!=\!\varepsilon_0\sqrt{\left(1+\frac{\Omega^\pm_{\vec{k}}}{2}\right)\left[1-\Omega^\pm_{\vec{k}}\left(1-\frac{3B^2}{2B_0^2}\right)\right]}\!-\!\mu^z_{\pm1,\vec{k}}B,\\
    &\varepsilon_{0,\vec{k}}=\varepsilon_0\sqrt{\left(1-\Omega^c_{\vec{k}}\right)\left[1+\Omega^c_{\vec{k}}\left(2-3B^2/B_0^2\right)\right]}\!-\!\mu^z_{0,\vec{k}}B.
\end{align}
Here, the characteristic energy $\varepsilon_0=3J|g|\mu_{\textsc{b}}/\mu_s$ determines the energy scale for the magnons. We introduce the notations $\Omega^\pm_{\vec{k}}=\Omega^c_{\vec{k}}\pm\sqrt{3}\Omega^s_{\vec{k}}$, where $\Omega^c_{\vec{k}}=\frac13\sum_{n}\cos(\vec{k}\cdot\vec{\delta}_n)$, $\Omega^s_{\vec{k}}=\frac13\sum_{n}\sin(\vec{k}\cdot\vec{\delta}_n)$. Here $\vec{\delta}_1=\vec{a}_1$, $\vec{\delta}_2=\vec{a}_2-\vec{a}_1$, $\vec{\delta}_3=-\vec{a}_2$ is the auxiliary basis of the triangular lattice, see Fig.~\ref{fig:lattice}. In Eq.~\eqref{eq:E_k}, we introduce the magnon magnetic moments according to the definition in Eq.~\eqref{eq:mu}:
\begin{equation}\label{eq:mu_k}
    \mu^z_{\pm1,\vec{k}}=\frac{|g|\mu_{\textsc{b}}}{2}\left(\mp\,\Omega^c_{\vec{k}}+\frac{\Omega^s_{\vec{k}}}{\sqrt{3}}\right),\quad\mu^z_{0,\vec{k}}=-|g|\mu_{\textsc{b}}\frac{\Omega^s_{\vec{k}}}{\sqrt{3}}.
\end{equation}
In the vicinity of the $\Gamma$-point, the magnon magnetic moments are approximately constants: $\mu^z_{\pm1,\vec{k}}=\mp|g|\mu_{\textsc{b}}/2+\mathcal{O}(k^2)$, and $\mu^z_{0,\vec{k}}=\mathcal{O}(k^3)$~\footnote{Note that due to nonanalyticity of the dispersions $\varepsilon_{\pm1,\vec{k}}$ in the $\Gamma$-point, magnetic moments $\mu_{\pm1,\vec{k}}^z$ are ill-defined in the $\Gamma$-point.}.

\begin{figure}
    \centering
    \includegraphics[width=\columnwidth]{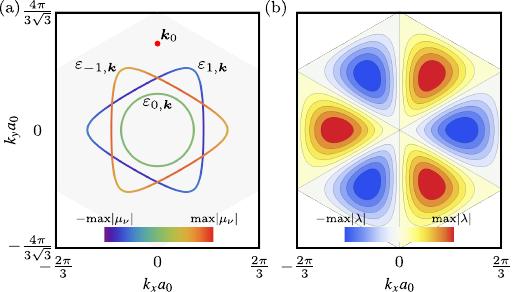}
    \caption{(a) -- The isosurfaces of constant energy $\varepsilon_{\nu,\vec{k}}=\varepsilon_0$ are built from the dispersion relations in Eq.~\eqref{eq:E_k} for the case $B=0$. The value of the magnon magnetic moment in Eq.~\eqref{eq:mu_k} is shown by the color scheme. The 1st Brillouin zone (1.BZ) corresponding to the magnetic unit cell is shown in gray shading. (b) -- Distribution of the spin-splitting parameter $\lambda$ defined in Eq.~\eqref{eq:lambda} over the 1.BZ demonstrates the $f$-wave character of the magnon spectra.}
    \label{fig:mu_lambda}
\end{figure}

By applying the external field within the spin plane, we find the corresponding magnon eigenenergies and show that the in-plane components of the magnon magnetic moment are absent, i.e. $\vec{\mu}_{\nu,\vec{k}}=\mu^z_{\nu,\vec{k}}\hat{\vec{z}}$, see Appendix~\ref{app:mu-ip}. This result is consistent with symmetry arguments: for the spins lying in the $xy$ plane, there is a $[C_{3z}\parallel E|\vec{t}]$ 
spin symmetry~\footnote{Here $C_{3z}$ denotes the 3-fold rotation in spin space, and $E$ is an identity operation.}, which suppresses the in-plane spin magnetic moment for any $\vec{k}$. (In spin symmetry notation~\cite{Litvin74}, the element to the left/right of the double bar acts exclusively in spin/real space.) 

Applying the symmetry properties $\Omega^c_{-\vec{k}}=\Omega^c_{\vec{k}}$, $\Omega^s_{-\vec{k}}=-\Omega^s_{\vec{k}}$, and $\Omega^\pm_{-\vec{k}}=\Omega^\mp_{\vec{k}}$ to \eqref{eq:mu_k} and \eqref{eq:E_k} for the case $B=0$, we obtain the previously discussed symmetry relations \eqref{eq:sym}. This symmetry is reflected in Fig.~\ref{fig:mu_lambda}(a), where the energy isolines and the corresponding values of the magnon magnetic moment are shown within the 1st Brillouin zone (1.BZ). In order to demonstrate the $f$-wave character explicitly, we compute the spin-splitting parameter~\cite{Kravchuk25a} 
\begin{equation}\label{eq:lambda}
   \lambda_{\vec{k}}=\frac
   {1}{|g|\mu_{\textsc{b}}\varepsilon_0}\sum\limits_\nu\mu^z_{\nu,\vec{k}}\varepsilon_{\nu,\vec{k}}
\end{equation}
and show its distribution over the 1.BZ in Fig.~\ref{fig:mu_lambda}(b). On both panels of Fig.~\ref{fig:mu_lambda}, we show the 1.BZ, which corresponds to the magnetic unit cell.  The latter contains three spins and is determined by the basis vectors $\vec{e}_1=2\vec{a}_1-\vec{a}_2$ and $\vec{e}_2=\vec{a}_1+\vec{a}_2$, see Fig.~\ref{fig:lattice}.  

\begin{figure}
    \centering
    \includegraphics[width=0.95\columnwidth]{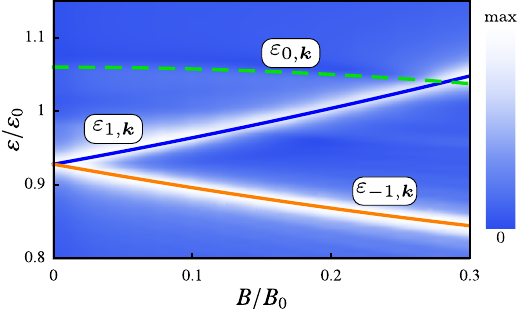}
    \caption{The effect of the Zeeman splitting of modes $\varepsilon_{\pm1,\vec{k}_0}(B)$ for the fixed wave vector $\vec{k}_0$ shown by the red dot in Fig.~\ref{fig:mu_lambda}(a). Here, lines show the eigen-energies~\eqref{eq:E_k} while the magnon intensities obtained by means of the spin lattice simulations (see Appendix~\ref{app:simuls} for the details) are shown by the color scheme.}
    \label{fig:Zeeman}
\end{figure}

Having a magnetic moment, magnons are subject to the Zeeman-splitting effect, which is demonstrated for the magnons $\varepsilon_{\pm1,\vec{k}_0}$ in Fig.~\ref{fig:Zeeman}. We choose the fixed wave vector $\vec{k}_0$ such that the  modes $\varepsilon_{\pm1,\vec{k}_0}$ are degenerate for zero field and have opposite magnetic moments $\mu^z_{1,\vec{k}_0}=-\mu^z_{-1,\vec{k}_0}$. In the small magnetic field $B$, these modes exhibit an energy splitting that is linear in $B$. The latter is also confirmed by the full-scale spin-lattice simulations, see Appendix~\ref{app:simuls} for details.

\begin{figure*}
    \centering
    \includegraphics[width=\textwidth]{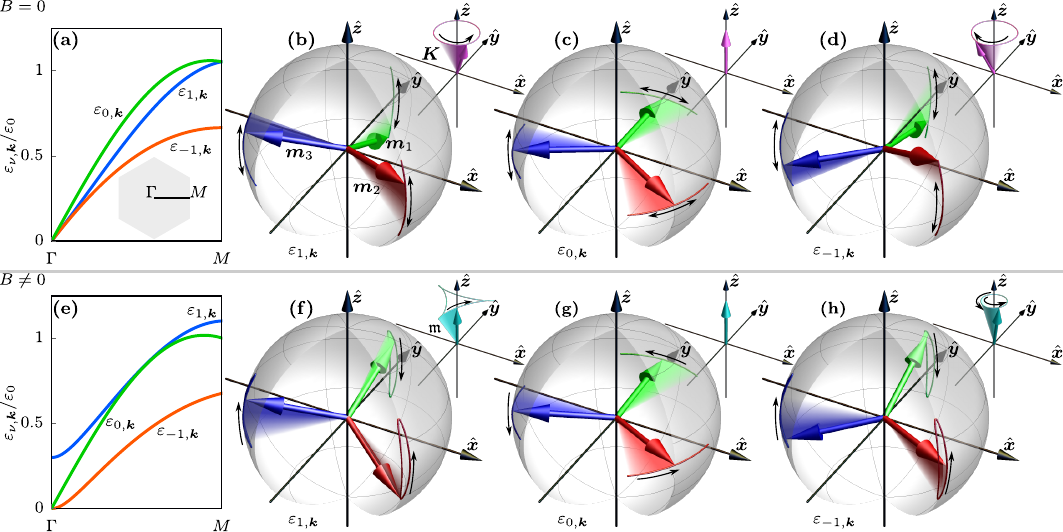}
    \caption{The influence of the applied magnetic field on the magnon modes is shown in terms of the dispersion relations, see panels (a) and (e), and in terms of the reconstructed single spin dynamics, see panels (b-d) and (f-h). The green, red, and blue arrows show three neighboring magnetic moments $\vec{m}_1$, $\vec{m}_2$, and $\vec{m}_3$ belonging to three different sublattices. The ground 120$\degree$ state lies in $xy$-plane. The insets show the dynamics of the vector spin chirality $\vec{K}$ and the total magnetization $\vec{\mathfrak{m}}$ for the cases without (upper row) and with (bottom row) an applied magnetic field, respectively. The applied field is $\vec{B}=0.1B_0\hat{\vec{z}}$. For panels (b-d) and (f-h), the wave vector $\vec{k}=10^{-3}\hat{\vec{x}}/a_0$ is fixed and close to the $\Gamma$-point.}
    \label{fig:m_reconstr}
\end{figure*}

For the case $B=0$, in the vicinity of $\Gamma$-point, all three magnon modes are gapless and linear: $\varepsilon_{\nu,\vec{k}}\approx\hbar u_{\nu}|\vec{k}|$, where $u_{\pm1}=(3/2)^{3/2}a_0J\gamma/\mu_s$ and $u_0=\sqrt{2}u_{\pm1}$ are the magnon velocities, see Fig.~\ref{fig:m_reconstr}(a). 
The Goldstone character of all three modes is due to the fact that the 120$\degree$ ground state spontaneously breaks continuous $SO(3)$ symmetry related to the arbitrariness of the orientation of the spin plane (modes $\varepsilon_{\pm1,\vec{k}}$) and continuous $SO(2)$ symmetry related to the arbitrariness of the orientation of the 120$\degree$-state within the spin plane (mode $\varepsilon_{0,\vec{k}}$). 
The small magnetic field ($0<B\ll B_0$) applied perpendicular to the spin plane acts on the magnon modes in markedly different ways, depending on their magnetic moments. While the mode $\varepsilon_{0,\vec{k}}$ with the vanishing magnetic moment remains unchanged in the vicinity of the $\Gamma$-point, the polarized modes $\varepsilon_{\pm1,\vec{k}}$ change drastically. For $\vec{k}=\vec{0}$, the dispersions are $\varepsilon_{\pm1,\vec{0}}=\frac{|g|\mu_{\textsc{b}}}{2}(|B|\pm B)$, i.e., the mode $\varepsilon_{1,\vec{k}}$ with the magnetic moment opposite to the field obtains a gap, while the mode $\varepsilon_{-1,\vec{k}}$ remains gapless. In the vicinity of $\Gamma$-point, these two modes become quadratic~\footnote{Note, that this approximation is not valid if $B=0$.}: $\varepsilon_{1,\vec{k}}\approx|g|\mu_{\textsc{b}}B+\hbar^2k^2/(2m^*)$, and $\varepsilon_{-1,\vec{k}}\approx\hbar^2k^2/(2m^*)$ with $m^*=4\hbar^2(B/B_0)/(\varepsilon_0a_0^2)$, see Fig.~\ref{fig:m_reconstr}(e). The applied magnetic field breaks the $SO(3)$ symmetry of the Heisenberg Hamiltonian, and therefore, the modes $\varepsilon_{\pm1,\vec{k}}$ are not Goldstone modes in the field.  The gaplessness of the mode $\varepsilon_{-1,\vec{k}}$ is accidental. In the next chapter, we demonstrate that this mode acquires a gap for a 3D extension of the model when an additional inter-layer exchange coupling is taken into account.

Computing the eigen-vectors for the magnon modes and using the Holstein-Primakoff representation, we reconstructed the single-spin dynamics, see Fig.~\ref{fig:m_reconstr}(b-d,f-h), and Appendix~\ref{app:spectra} for details. We chose a wave vector very close to the $\Gamma$-point. In this case, the trihedron $\{\vec{m}_1,\vec{m}_2,\vec{m}_3\}$ formed by the neighboring magnetic moments belonging to different sublattices exhibits a rigid-body dynamics, such that all three vectors stay in one plane. Each magnetic moment exhibits linearly polarized oscillations within the ground-state plane (mode $\varepsilon_{0,\vec{k}}$) or perpendicular to it (modes $\varepsilon_{\pm1,\vec{k}}$). The difference between modes $\varepsilon_{1,\vec{k}}$ and $\varepsilon_{-1,\vec{k}}$ is the opposite directions of the trihedron wobbling, which coincides with the direction of precession of the vector spin chirality~\cite{Kawamura84} $\vec{K}=\vec{m}_1\times\vec{m}_2+\vec{m}_2\times\vec{m}_3+\vec{m}_3\times\vec{m}_1$, see the insets in Figs.~\ref{fig:m_reconstr}(b-d). The mechanics of the described low-energy modes for the case $B=0$ has been identified in a number of previous works; see Refs.~\cite{Pradenas25,Adamyan26}.

The applied magnetic field does not qualitatively change the dynamics of the unpolarized mode $\varepsilon_{0,\vec{k}}$, see Fig.~\ref{fig:m_reconstr}(g). However, it significantly changes the dynamics of the polarized modes $\varepsilon_{\pm1,\vec{k}}$. Magnetic moments $\vec{m}_i$ exhibit elliptical precession, with the major axis of the ellipse aligned with the field. Although all $\vec{m}_i$ in both modes precess in the same direction, due to the different phase shifts between the magnetic moments, vector $\vec{K}$ precesses in opposite directions for modes $\varepsilon_{1,\vec{k}}$ and $\varepsilon_{-1,\vec{k}}$, and its dynamics is similar to the case $B=0$. The total magnetic moment vector $\vec{\mathfrak{m}}=\sum_i\vec{m}_i$, however, precesses in the same direction for both modes; see Fig.~\ref{fig:m_reconstr}(f,h). One should note the following unusual properties of the mode $\varepsilon_{-1,\vec{k}}$ whose magnetic moment is directed along the magnetic field: (i) vectors $\vec{K}$ and $\vec{\mathfrak{m}}$ precess in opposite directions, (ii) vector $\vec{\mathfrak{m}}$ makes two rotations per period, (iii) the trihedron $\{\vec{m}_1,\vec{m}_2,\vec{m}_3\}$ does not exhibit the rigid-body dynamics, (iv) it is a gapless non-Goldstone mode.

\section{Multilayer system -- Magnons and transport properties.}\label{sec:mult}
\begin{figure}
    \centering
\includegraphics[width=\columnwidth]{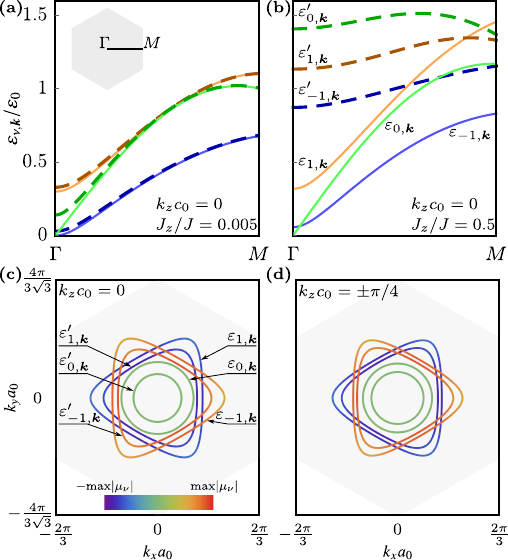}
    \caption{The influence of the interlayer coupling on the magnon spectra is shown in panels (a,b). The isosurfaces of the constant energy $\varepsilon_{\nu,\vec{k}}=0.6\varepsilon_0$ for all six modes for the case $J_z/J=0.05$ are shown in panels (c,d) for different values of $k_z$. The isosurface color shows the magnon magnetic moment. Magnetic field is $B=0.1B_0$ and $B=0$ for panels (a,b), and (c,d), respectively.}
    \label{fig:spct_3D}
\end{figure}

Here, we introduce a 3D generalization of our model as an infinite stack of the monolayers considered in the previous section. The layers are coupled by the antiferromagnetic exchange interaction of strength $J_z$. In the ground state, the two magnetic moments, one above the other in adjacent layers, point in opposite directions (in the absence of a magnetic field).
Thus, the number of sublattices doubles. 
As a consequence, we obtain three additional magnon modes, $\varepsilon'_{\pm1,\vec{k}}$, and $\varepsilon'_{0,\vec{k}}$, so six in total, see Fig.~\ref{fig:spct_3D}.
Magnetic field applied perpendicularly to the spin plane forms the ``umbrella'' state; however, with a smaller canting $m_z=\varsigma B/B_0$, where $\varsigma=(1+\frac49\frac{J_z}{J})^{-1}<1$. Modes $\varepsilon'_{\nu,\vec{k}}$ and $\varepsilon_{\nu,\vec{k}}$ have the same magnetic moments, i.e. $\bar{\mu}'^z_{\nu,\vec{k}}=\bar{\mu}^z_{\nu,\vec{k}}=\varsigma \mu^z_{\nu,\vec{k}}$, see Appendix~\ref{app:3D} for details. Remarkably, due to the scaling factor $\varsigma$, the magnon magnetic moment $\bar{\mu}^z_{\nu,\vec{k}}$ vanishes in the limit $J_z\gg J$. Easy-plane anisotropy leads to a similar suppression of the magnon magnetic moment. In both cases, the suppression occurs due to an increase in the stiffness of the magnetic subsystem with respect to the applied magnetic field.

In the limit $J_z\to0$, the modes become doubly degenerate, i.e., $\varepsilon'_{\nu,\vec{k}}\to\varepsilon_{\nu,\vec{k}}$, see Fig.~\ref{fig:spct_3D}(a). However, the finite $J_z$ gaps out the new modes, turning them to the form $\varepsilon'_{\nu,\vec{k}}=\Delta'_{\nu}+\mathcal{O}(k^2_i)$ in the vicinity of the $\Gamma$-point. In the limit $J_z/J\ll1$ and $B/B_0\ll1$, the gaps are $\Delta'_0\approx2\varepsilon_0\sqrt{
J_z/J}$ and  $\Delta'_{\pm1}\approx\frac32\varepsilon_0[\sqrt{B^2/B^2_0+8J_z/(9J)}\pm B/B_0]$, see Fig.~\ref{fig:spct_3D}(b). Thus, the gaps do not vanish even for a vanishing magnetic field. 

For the case $B=0$, in the limit $J_z/J\ll1$, the Goldstone modes are trivially modified due to the presence of $k_z$-vector: $\varepsilon_{\nu,\vec{k}}\approx\hbar\sqrt{u^2_\nu(k^2_x+k^2_y)+u_{\nu,z}^2k_z^2}$, where $u_{0,z}=3c_0\gamma\sqrt{J_zJ}/\mu_s$ and $u_{\pm1,z}=u_{0,z}/\sqrt{2}$ are the vertical components of the magnon velocity, with $c_0$ being the inter-layer spacing. For the case $B>0$, the mode $\varepsilon_{0,\vec{k}}$ is the only Goldstone mode. Being unpolarized (in the $\Gamma$-point), it experiences a negligible effect of the applied small magnetic field, while the polarized modes are gapped in the field: $\varepsilon_{\nu,\vec{k}}=\Delta_{\nu}+\mathcal{O}(k_i^2)$.
In the limit $J_z/J\ll1$ and $B/B_0\ll1$, the mode $\varepsilon_{1,\vec{k}}$, which is polarized against the applied field, has the same gap $\Delta_1\approx|g|\mu_{\textsc{b}}B$ as for the monolayer case, while the mode $\varepsilon_{-1,\vec{k}}$, which is gapless for the monolayer, in 3D, obtains gap $\Delta_{-1}\approx\frac49|g|\mu_{\textsc{b}}B J_z/J$. The latter shows that the gaplessness of this mode for a monolayer is accidental. 

Since the Mermin-Wagner theorem does not forbid magnetic ordering at sufficiently low finite temperatures in 3D systems, we consider in the following two magnon transport effects caused by a temperature gradient.

\subsection{Magnon spin-splitter effect}\label{sec:spin-splitter}
\begin{figure}
    \centering
    \includegraphics[width=\columnwidth]{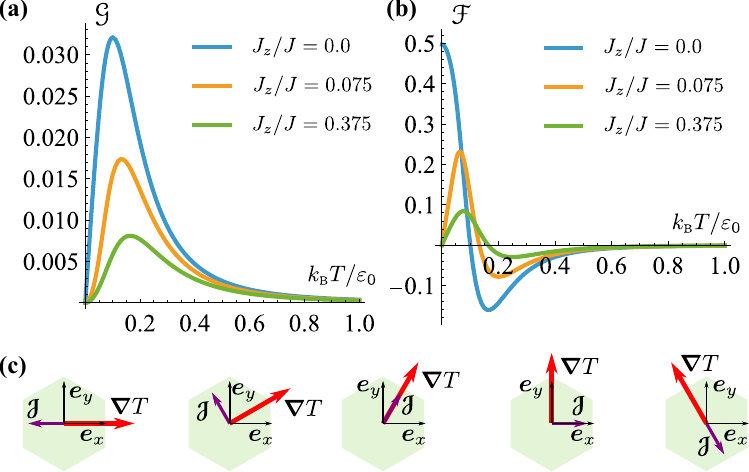}
    \caption{The temperature dependencies of the dimensionless strengths of the nonlinear spin-splitter and Edelstein effects are shown in panels (a) and (b), respectively. Functions $\mathcal{G}$ and $\mathcal{F}$ are introduced in Eqs.~\eqref{eq:G2} and \eqref{eq:F3}, respectively. Panel (c) demonstrates the dependence of the orientation of the thermal spin current $\vec{\mathcal{J}}$ on the orientation of the temperature gradient relative to the Brillouin zone.}
    \label{fig:GF-vs-T}
\end{figure}

Here, we consider the magnon-driven spin current (the flow of magnetic moment) induced by a temperature gradient for the case $\vec{B}=0$. We start from the general quasi-classical expression for the spin current 
\begin{equation}\label{eq:J-gen}
    \vec{\mathcal{J}}=\frac{1}{V}\sum\limits_\nu\sum\limits_{\vec{k}\in\text{1.BZ}}\bar{\mu}^z_{\nu,\vec{k}}\vec{v}_{\nu,\vec{k}}\,\delta n_{\nu,\vec{k}},
\end{equation}
where $V$ is the sample volume, $\nu$ numerates all six magnon branches, $\vec{k}$ runs over the 3D Brillouin zone, $\vec{v}_{\nu,\vec{k}}=\hbar^{-1}\partial_{\vec{k}}\varepsilon_{\nu,\vec{k}}$ is the group velocity, and $\delta n_{\nu,\vec{k}}$ denotes the nonequilibrium part of the distribution function, caused by the coordinate dependence of temperature. Using the Boltzmann transport equation in the relaxation-time approximation, for the case of constant temperature gradients~\footnote{I.e, it is assumed that $\partial^2_{\alpha\beta}T=0$ and the same for the higher derivatives.}, one can write
\begin{equation}\label{eq:J}
    \mathcal{J}_\alpha=G_{\alpha\beta}\partial_{\beta}T+G^{(2)}_{\alpha\beta\gamma}\partial_{\beta}T\partial_{\gamma}T+\dots,
\end{equation}
where $\alpha,\beta,\gamma\in\{x,y,z\}$. For details, see Appendix~\ref{app:transport}. While the linear thermal spin conductivity vanishes $G_{\alpha\beta}=0$ for the considered model~\cite{Ezawa25}, the non-zero components of its nonlinear counterpart are related as $-G_{xxx}^{(2)}=G_{xyy}^{(2)}=G_{yxy}^{(2)}=G_{yyx}^{(2)}\equiv G^{(2)}$, where
\begin{equation}\label{eq:G2}
    G^{(2)}=\frac{\tau^2_{\text{rlx}}k^2_{\textsc{b}}\varepsilon_0a_0|g|\mu_{\textsc{b}}}{c_0\hbar^3}\mathcal{G}\left(\frac{k_{\textsc{b}}T}{\varepsilon_0};\frac{J_z}{J}\right),
\end{equation}
and the dimensionless function $\mathcal{G}$ is shown in Fig.~\ref{fig:GF-vs-T}(a). Here, the relaxation time $\tau_{\mathrm{rlx}}$ is the average time between two magnon collisions. Note that the obtained nonlinear spin-conductivity tensor $G^{(2)}_{\alpha\beta\gamma}$ exhibits the same symmetry features as those recently predicted for the electron spin currents in $f$-wave magnets~\cite{Ezawa25,Ezawa26}. In the low-temperature limit $k_{\textsc{b}}T\ll\varepsilon_0\sqrt{J_z/J}$ and for the case $J_z\ll J$ we estimate
\begin{equation}\label{eq:G-cal}
    \mathcal{G}\left(\frac{k_{\textsc{b}}T}{\varepsilon_0};\frac{J_z}{J}\right)\approx\frac{6\sqrt{6}\pi^2}{175}\sqrt{\frac{J}{J_z}}\left(\frac{k_{\textsc{b}}T}{\varepsilon_0}\right)^2.
\end{equation}

Since the application of the temperature gradient in the $z$-direction does not produce the spin current, in the following we consider only in-plane gradients $\vec{\nabla}T=|\vec{\nabla}T|(\hat{\vec{x}}\cos\chi+\hat{\vec{y}}\sin\chi)$. In this case 
\begin{equation}\label{eq:J-dT}
    \vec{\mathcal{J}}=G^{(2)}|\vec{\nabla}T|^2(-\vec{e}_x\cos2\chi+\vec{e}_y\sin2\chi).
\end{equation}
Thus, the relative orientation of $\vec{\mathcal{J}}$ and $\vec{\nabla}T$ is not fixed, it depends on how $\vec{\nabla}T$ is applied relative to the lattice directions, see Fig.~\ref{fig:GF-vs-T}(c). Namely, the orientation of $\vec{\nabla}T$ in one of the six directions towards a nearest (next-nearest) neighbor corresponds to the longitudinal (perpendicular) relative orientations of $\vec{\nabla}T$ and $\vec{\mathcal{J}}$.  

Since the magnon spin (angular) momentum~\cite{Okuma17} is proportional to the magnon magnetic moment~\cite{Neumann20}, the same experimental setup used to measure the thermally-induced spin currents in insulating compensated magnets~\cite{Yang26b,Wu16a} is well suited for the measurement of $\mathcal{J}$. 

\subsection{Magnon Edelstein effect}

Here, we consider the effect of the system's total magnetic moment generation in response to the applied temperature gradient. The magnetic moment is directed in the $z$-direction, and its density is
\begin{equation}\label{eq:M}
    M=\frac{1}{V}\sum\limits_{\nu}\sum\limits_{\vec{k}\in\text{1.BZ}}\bar{\mu}^z_{\nu,\vec{k}}\,\delta n_{\nu,\vec{k}}.
\end{equation}
Here, as previously, we consider the summation over six magnon branches in the 3D Brillouin zone. The contribution to $M$ from the equilibrium part of the distribution function is averaged out because of the symmetry of the distribution of $\bar{\mu}^z_{\nu,\vec{k}}$ over the magnon branches. With the help of the Boltzmann equation, we obtain
\begin{equation}\label{eq:M-exp}
    M=F_\alpha\,\partial_\alpha T+F^{(2)}_{\alpha\beta}\,\partial_{\alpha}T\,\partial_\beta T+F^{(3)}_{\alpha\beta\gamma}\,\partial_\alpha T\,\partial_\beta T\,\partial_\gamma T+\dots,
\end{equation}
where, as previously, we assumed the constant temperature gradients, see Appendix~\ref{app:transport}. We find that $F_{\alpha}=0$ and $F_{\alpha\beta}=0$, and the leading-order contribution comes from the 3rd-order nonlinear terms, which have the following symmetry $F_{xxx}^{(3)}=-F_{xyy}^{(3)}=-F_{yxy}^{(3)}=-F_{yyx}^{(3)}\equiv F^{(3)}$. Here 
\begin{equation}\label{eq:F3}
    F^{(3)}=\frac{\tau^3_{\text{rlx}}k^3_{\textsc{b}}a_0|g|\mu_{\textsc{b}}}{c_0\hbar^3}\mathcal{F}\left(\frac{k_{\textsc{b}}T}{\varepsilon_0};\frac{J_z}{J}\right),
\end{equation}
where function $\mathcal{F}$ is presented in Fig.~\ref{fig:GF-vs-T}(b). One can show that $\mathcal{F}(x,y)=\partial_{x}\mathcal{G}(x,y)$, see Appendix~\ref{app:transport} for the details. In the low-temperature limit $k_{\textsc{b}}T\ll\varepsilon_0\sqrt{J_z/J}$ and for the case $J_z\ll J$ we estimate
\begin{equation}\label{eq:F-cal}
    \mathcal{F}\left(\frac{k_{\textsc{b}}T}{\varepsilon_0};\frac{J_z}{J}\right)\approx\frac{12\sqrt{6}\pi^2}{175}\sqrt{\frac{J}{J_z}}\frac{k_{\textsc{b}}T}{\varepsilon_0}.
\end{equation}
Taking into account the symmetry of tensor $F^{(3)}_{\alpha\beta\gamma}$, we derive
\begin{equation}\label{eq:M-Edel}
    M=F^{(3)}|\vec{\nabla}T|^3\cos3\chi
\end{equation}
                    meaning that the extrema values of $M$ appear when $\vec{\nabla}T$ is applied in the directions of maximum spin splitting.

Let's estimate the order of magnitude of the maximal magnetization $M_{\text{max}}$ expected in an experiment.
Assuming that $a_0\approx c_0$, and using the prefactor in Eq.~\eqref{eq:F3} as a figure of merit to estimate $F^{(3)}$, we obtain
\begin{equation}\label{eq:Mmax_est}
    M_{\text{max}}\sim\frac{\mu_{\textsc{b}}|g|}{\Delta L^3}\left(\frac{k_{\textsc{b}}\Delta T}{\Gamma_{\vec{k}}}\right)^3,
\end{equation}
where $\Gamma_{\vec{k}}=\hbar/\tau_{\text{rlx}}$ is the decay rate and we denote $|\vec{\nabla}T|=\Delta T/\Delta L$. Based on the data of inelastic neutron scattering for metallic CrB$_2$~\cite{Park20}, we estimate $\langle\Gamma_{\vec{k}}\rangle\sim0.5$~meV as the integral averaged~\footnote{We used the fitting function $\Gamma_{\vec{k}}\approx ck^5$, which corresponds the 3D case~\cite{Chernyshev09}.} for $|\vec{k}|a_0<0.1$~\footnote{For $\varepsilon_0\sim10$~meV~\cite{Park20}, it corresponds to $T<0.13\varepsilon_0/k_{\textsc{b}}\approx15$~K. Note that $T_N=88$~K for CrB$_2$~\cite{Park20}.}.  The corresponding relaxation time is $\tau_{\text{rlx}}\approx\hbar/\langle\Gamma_{\vec{k}}\rangle\sim10^{-12}$~s. For the insulators, it is reasonable to assume that the relaxation time is at least two orders of magnitude larger, i.e., $\tau^{\text{ins}}_{\text{rlx}}\sim10^{-10}$~s. For the estimation, we use the temperature gradient value $|\vec{\nabla}T|\approx0.01$~K/nm achievable for transport experiments~\cite{Bose18}. Note that this value satisfies the limitation $|\vec{\nabla}T|\ll\langle\Gamma^{\text{ins}}_{\vec{k}}\rangle/(a_0k_{\textsc{b}})\approx0.05$~K/$a_0$ required for the performed perturbative treatment of the Boltzmann equation. The magnetization estimated in \eqref{eq:Mmax_est} is $M_{\text{max}}\sim10^{-8}\,\mu_{\textsc{b}}/\text{nm}^3$ and $M_{\text{max}}\sim10^{-2}\,\mu_{\textsc{b}}/\text{nm}^3$ for the cases of a metal and an insulator, respectively. While the Edelstein magnetization is hardly measurable for metals, for insulators it can be detected by NV-center diamond magnetometry~\cite{Finco23}.
                    

\section{Conclusions}
We find that the magnons excited on the top of the frustrated 120$\degree$ state of the triangular lattice carry a magnetic moment directed perpendicular to the spin plane. The appearance of this magnon magnetic moment is counterintuitive, since none of the sublattices is magnetized in the perpendicular direction. Recently, it was shown~\cite{Pradenas25} that the magnons in this system possess an additional quantum number \emph{isospin}, beyond energy and momentum. The relationship between the magnon magnetic moment and the isospin will be analyzed in detail in a subsequent paper. 

We predict measurable thermal transport effects that rely on the found magnon magnetic moment, namely the nonlinear magnon spin-splitter and magnon Edelstein effects. The value of the net magnetization, which is generated due to the Edelstein effect in response to the applied temperature gradient, is estimated to be in the range accessible for the experimental observations.

The bond-dependent anisotropy~\cite{Maksimov19}, which is present in realistic materials, introduces an additional source of frustration and its influence on the magnon magnetic moment should be considered. This influence can be especially large in the regimes close to the phase transition between the 120$\degree$ and stripe states~\cite{Maksimov19}. The latter transition can also take place under the combined action of the uniaxial anisotropy and interfacial Dzyaloshinskii-Moriya interaction~\cite{Fang21} or the next-nearest-neighbor exchange interaction~\cite{Gallegos25}. The additional next-nearest-neighbor exchange couplings can also lead to the formation of the spin-liquid phase~\cite{Gallegos25,Glittum26}. The study of the evolution of the magnon magnetic moment in these regimes and the related transport properties constitutes the direction of further development of this project.

\section{Acknowledgments}
V.K.~acknowledges financial support from the Deutsche
Forschungsgemeinschaft (DFG, German Research Foundation) through the W{\"u}rzburg-Dresden Cluster of Excellence ctd.qmat – Complexity, Topology and Dynamics
in Quantum Matter (EXC 2147, project-id 390858490). K.Y. ~acknowledges financial support from the Deutsche Forschungsgemeinschaft (DFG, German Research Foundation) through grant No. YE 232/2-1. B.P. acknowledges support from the Leibniz Association through the Leibniz Competition Project No. J200/2024.

The work of R.R.N.~was funded by the German Research Foundation (DFG) as part of the German Excellence Strategy EXC3112/1-533767171 (Center for Chiral Electronics) and through TRR 277-328545488 (Project No.~B04). The work of A.M.~was funded by the German Research Foundation (DFG) as part of TRR 173-268565370 (Project No.~B13) and Project No.~504261060 (Emmy Noether Programme). The work of J.S.~was funded by the German Research Foundation (DFG) as part of TRR 173-268565370 (Project No. ~03), and TRR 288-422213477 (Projects No.~A09 and B05). A.M.~and J.S.~acknowledge support by the Dynamics and Topology Center (TopDyn) funded by the State of Rhineland-Palatinate.

We thank Ulrike Nitzsche for the technical support.

\appendix

\section{Spectrum of spin waves for the single layer}\label{app:spectra}

We consider a triangular lattice of spins shown in Fig.~\ref{fig:lattice}. The antiferromagnetic Heisenberg exchange with $J>0$ acts between the nearest neighbors, and a magnetic field $\vec{B}$ is applied perpendicularly to the lattice. The discrete position vector $\vec{R}_{\vec{n}}=n_1\vec{e}_1+n_2\vec{e}_2$ with $n_1,\;n_2\in\mathbb
{Z}$ numerate \emph{magnetic} unit cells determined by the basis vectors $\vec{e}_1=\vec{\delta}_1-\vec{\delta}_2$ and $\vec{e}_2=\vec{\delta}_1-\vec{\delta}_3$, see the orange shadowing in Fig.~\ref{fig:lattice}. For the convenience of subsequent calculations, it is convenient to write Hamiltonian~\eqref{eq:H} as a sum over $\vec{R}_{\vec{n}}$:
\begin{equation}\label{eq:H-det}
    \begin{split}
&\mathcal{H}=J\sum\limits_{\vec{R}_{\vec{n}}}\Bigl\{\\
&\vec{m}_1(\vec{R}_{\vec{n}})\!\cdot\!\!\bigl[\vec{m}_2(\vec{R}_{\vec{n}}\!+\!\vec{\delta}_1)+\vec{m}_2(\vec{R}_{\vec{n}}\!+\!\vec{\delta}_2)+\vec{m}_2(\vec{R}_{\vec{n}}\!+\!\vec{\delta}_3)\big]\\
    +&\vec{m}_2(\vec{R}_{\vec{n}})\!\cdot\!\bigl[\vec{m}_3(\vec{R}_{\vec{n}}\!+\!\vec{\delta}_1)+\vec{m}_3(\vec{R}_{\vec{n}}\!+\!\vec{\delta}_2)+\vec{m}_3(\vec{R}_{\vec{n}}\!+\!\vec{\delta}_3)\bigr]\\
    +&\vec{m}_3(\vec{R}_{\vec{n}})\!\cdot\!\bigl[\vec{m}_1(\vec{R}_{\vec{n}}\!+\!\vec{\delta}_1)\!+\!\vec{m}_1(\vec{R}_{\vec{n}}\!+\!\vec{\delta}_2)\!+\!\vec{m}_1(\vec{R}_{\vec{n}}\!+\!\vec{\delta}_3)\bigr]\Bigr\}\\
    &-B\mu_s\sum\limits_{\vec{R}_{\vec{n}}}\bigl[m_{1z}(\vec{R}_{\vec{n}})\!+\!m_{2z}(\vec{R}_{\vec{n}}\!+\!\vec{\delta}_1)\!+\!m_{3z}(\vec{R}_{\vec{n}}\!+\!2\vec{\delta}_1)\bigr].
    \end{split}
\end{equation}
When writing Hamiltonian \eqref{eq:H}, we used the property $\sum_{\vec{R}_{\vec{n}}}\vec{m}_{\ell_1}(\vec{R}_{\vec{n}}+\vec{\delta R}_1)\cdot \vec{m}_{\ell_2}(\vec{R}_{\vec{n}}+\vec{\delta R}_2)=\sum_{\vec{R}_{\vec{n}}}\vec{m}_{\ell_1}(\vec{R}_{\vec{n}})\cdot \vec{m}_{\ell_2}(\vec{R}_{\vec{n}}+\vec{\delta R}_2-\vec{\delta R}_1)$ which is equivalent to redefinition of the summation variable $\vec{R}_{\vec{n}}$.

The ground state of Hamiltonian \eqref{eq:H-det} is
\begin{equation}\label{eq:grnd}
    \begin{split}
    &\vec{m}^0_\ell=\sin\Theta\left[\vec{e}_x\cos\Phi_\ell+\vec{e}_y\sin\Phi_\ell\right]+\vec{e}_z\cos\Theta,\\
    &\Phi_\ell=\frac{2\pi(\ell-1)}{3}+\varphi_0,\qquad\cos\Theta=\frac{B\mu_s}{9J},
    \end{split}
\end{equation}
where $\ell=1,\,2,\,3$ and $\varphi_0$ is an arbitrary phase. Fig.~\ref{fig:lattice} shows a particular case $\varphi_0=\pi/2$. Deviations of magnetic moments $\vec{m}_\ell$ from their ground states $\vec{m}^0_\ell$ are described by the complex-valued functions $\psi_\ell$ defined by means of the classical form of the Holstein-Primakoff representation
\begin{align}\label{eq:HP}
    \nonumber\vec{m}_\ell=&\left(1-\frac{|\psi_\nu|^2}{S}\right)\vec{m}^0_\ell\\
    &+\sqrt{2-\frac{|\psi_\nu|^2}{S}}(\vec{e}^+_\ell\psi_\ell+\vec{e}^-_\ell\psi^*_\ell),\\
    \vec{e}^\pm_\ell=&\frac{1}{2\sqrt{S}}\bigl[\left(\cos\Theta\cos\Phi_\ell\mp i\sin\Phi_\ell\right)\vec{e}_x\\
    \nonumber&+\left(\cos\Theta\sin\Phi_\ell\pm i\cos\Phi_\ell\right)\vec{e}_y-\sin\Theta\vec{e}_z\bigr],
\end{align}
where $S=\mu_s/(|g|\mu_{\textsc{b}})$.
Here vectors $\vec{e}^\pm_\ell=\frac12(\vec{a}_\ell\pm i\vec{b}_\ell)/\sqrt{S}$ with $\vec{a}_\ell=\partial_\Theta\vec{m}^0_{\ell}$ and $\vec{b}_\ell=\vec{m}^0_\ell\times\vec{a}_\ell$ compose a basis in the plane perpendicular to $\vec{m}^0_\ell$. In terms of $\psi_\ell$, Landau-Lifshitz equation~\eqref{eq:LL} obtains the following form
\begin{equation}\label{eq:Psi}
    i\hbar\,\dot{\psi}_\ell(\vec{R}^\ell_{\vec{n}})=\frac{\partial\mathcal{H}}{\partial\psi^*_\ell(\vec{R}^\ell_{\vec{n}})}.
\end{equation}
By distributing all spins into the sublattices, we can identify each spin $\vec{m}_\ell$ using vector $\vec{R}^\ell_{\vec{n}}=\vec{R}_{\vec{n}}+(\ell-1)\vec{\delta}_1$ instead of $\vec{r}_{\vec{n}}$. 
Applying the Fourier transform 
\begin{equation}\label{eq:FT}
    \begin{split}
    &\psi_\ell(\vec{R}^\ell_{\vec{n}})=\frac{1}{\sqrt{N}}\sum\limits_{\vec{k}\in\text{1.BZ}}\hat{\psi}_\ell(\vec{k})e^{i\vec{R}^\ell_{\vec{n}}\cdot\vec{k}},\\
    &\hat{\psi}_\ell(\vec{k})=\frac{1}{\sqrt{N}}\sum\limits_{\vec{R}^\ell_{\vec{n}}}\psi_\ell(\vec{R}^\ell_{\vec{n}})e^{-i\vec{R}^\ell_{\vec{n}}\cdot\vec{k}}
    \end{split}
\end{equation}
supplemented with the completeness relation $\sum_{\vec{R}^\ell_{\vec{n}}}e^{i(\vec{k}-\vec{k}')\cdot\vec{R}^\ell_{\vec{n}}}=N\delta_{\vec{k},\vec{k}'}$ with $N$ being the number of the primitive magnetic unit cells, to the linearized (with respect to $\psi$) Eq.~\eqref{eq:Psi}, we obtain
\begin{equation}\label{eq:Psi-FT-lin}
    i\hbar\,\dot{\hat{\psi}}_\ell(\vec{k})=\frac{\partial\mathcal{H}^{(2)}}{\partial\hat{\psi}^*_\ell(\vec{k})}.
\end{equation}
Here $\mathcal{H}^{(2)}$ is harmonic (with respect to $\hat{\psi}_\ell$) part of $\mathcal{H}$. The substitution of \eqref{eq:HP} into \eqref{eq:H-det} with the subsequent application of the Fourier transform results in
\begin{align}\label{eq:H2}
    \nonumber&\mathcal{H}^{(2)}=\varepsilon_0\sum\limits_{\vec{k}\in\text{1.BZ}}\Bigl\{\frac12\left[|\hat{\psi}_1(\vec{k})|^2+|\hat{\psi}_2(\vec{k})|^2+|\hat{\psi}_3(\vec{k})|^2\right]\\
    &+\!\mathcal{A}\alpha_{\vec{k}}\left[\hat{\psi}_1(\vec{k})\hat{\psi}^*_2(\vec{k})\!+\!\hat{\psi}_2(\vec{k})\hat{\psi}^*_3(\vec{k})\!+\!\hat{\psi}_3(\vec{k})\hat{\psi}^*_1(\vec{k})\right]\\
    \nonumber&+\!\mathcal{B}\alpha_{\vec{k}}\!\left[\hat{\psi}_1(\vec{k})\hat{\psi}_2(\!-\vec{k})\!+\!\hat{\psi}_2(\vec{k})\hat{\psi}_3(\!-\vec{k})\!+\!\hat{\psi}_3(\vec{k})\hat{\psi}_1(\!-\vec{k})\right]\!\!+\!\text{c.c.}\!\Bigr\},
\end{align}
where $\mathcal{A}=\frac{1}{4}(1+i\sqrt{3}\cos\Theta)^2$, $\mathcal{B}=\frac34\sin^2\Theta$, and $\alpha_{\vec{k}}=(e^{-i\vec{k}\cdot\vec{\delta}_1}+e^{-i\vec{k}\cdot\vec{\delta}_2}+e^{-i\vec{k}\cdot\vec{\delta}_3})/3$.

\begin{figure*}
    \centering
    \includegraphics[width=\linewidth]{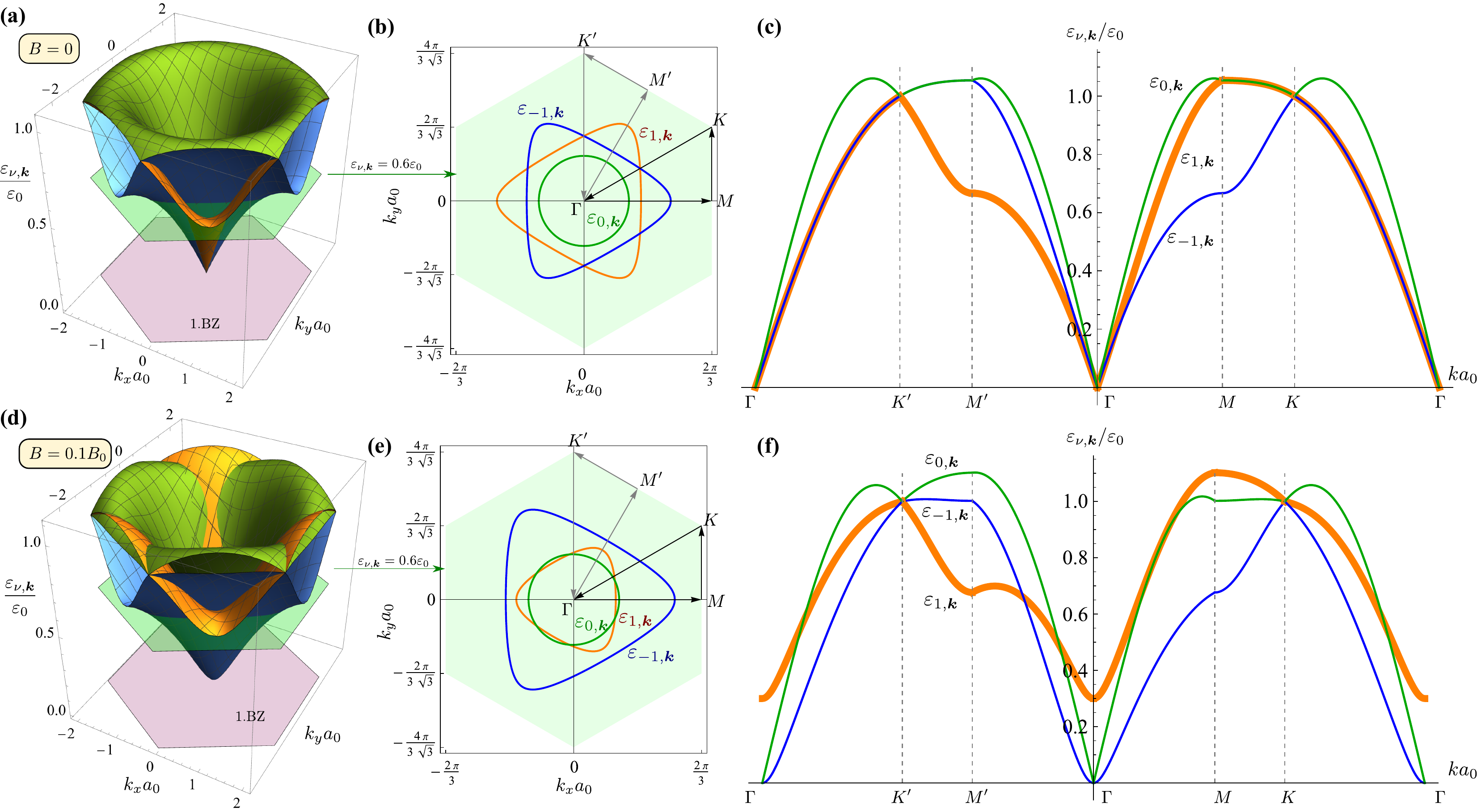}
    \caption{Dispersion relations \eqref{eq:E_k} are shown for $B=0$ (upper row) and for $B=0.1B_0$ (bottom row).}
    \label{fig:disp}
\end{figure*}

Introducing the Nambu spinor $\hat{\vec{\Psi}}_{\vec{k}}=[\hat{\psi}_1(\vec{k}),\hat{\psi}_2(\vec{k}),\hat{\psi}_3(\vec{k}),\hat{\psi}^*_1(-\vec{k}),\hat{\psi}^*_2(-\vec{k}),\hat{\psi}^*_3(-\vec{k})]^{T}$, we formulate Eqs.~\eqref{eq:Psi-FT-lin} in the form of a Schr{\"o}dinger equation
\begin{equation}\label{eq:Schrod}
    i\dot{\hat{\vec{\Psi}}}_{\vec{k}}=\varepsilon_0\eta\mathbb{H}_{\vec{k}}\hat{\vec{\Psi}}_{\vec{k}},
\end{equation}
where 
\begin{align}
    \nonumber\mathbb{H}_{\vec{k}}=\begin{bmatrix}
 \mathbb{A}_{\vec{k}} & \mathbb{B}_{\vec{k}} \\
        \mathbb{B}^*_{-\vec{k}} & \mathbb{A}^*_{-\vec{k}}
    \end{bmatrix},\quad\eta=\text{diag}(1,1,1,-1,-1,-1),\\
    \mathbb{A}_{\vec{k}}=\begin{bmatrix}
        1 & \mathcal{A}^*\alpha^*_{\vec{k}} & \mathcal{A}\alpha_{\vec{k}} \\
        \mathcal{A}\alpha_{\vec{k}} & 1 & \mathcal{A}^*\alpha^*_{\vec{k}} \\
        \mathcal{A}^*\alpha^*_{\vec{k}} & \mathcal{A}\alpha_{\vec{k}} & 1
    \end{bmatrix},\;\mathbb{B}_{\vec{k}}=\mathcal{B}\begin{bmatrix}
        0 & \alpha^*_{\vec{k}} & \alpha_{\vec{k}} \\
        \alpha_{\vec{k}} & 0 & \alpha^*_{\vec{k}} \\
        \alpha^*_{\vec{k}} & \alpha_{\vec{k}} & 0
    \end{bmatrix}.
\end{align}
Note that $\alpha^*_{-\vec{k}}=\alpha_{\vec{k}}$, therefore $\mathbb{B}^*_{-\vec{k}}=\mathbb{B}_{\vec{k}}$. Here $\mathbb{H}_{\vec{k}}$ is a Hermitian matrix which determines the harmonic part of the Hamiltonian $\mathcal{H}^{(2)}=\frac{\varepsilon_0}{2}\sum_{\vec{k}}\hat{\vec{\Psi}}^\dagger_{\vec{k}}\mathbb{H}_{\vec{k}}\hat{\vec{\Psi}}_{\vec{k}}$. 
The solution $\hat{\vec{\Psi}}_{\vec{k}}=\hat{\vec{\Psi}}_{\vec{k}}^0e^{-i\varepsilon t/\hbar}$ reduces Eq.~\eqref{eq:Schrod} to the eigenvalue problem (EVP)
\begin{equation}\label{eq:EVP}
    \frac{\varepsilon}{\varepsilon_0}\hat{\vec{\Psi}}_{\vec{k}}^0=\eta\mathbb{H}_{\vec{k}}
\hat{\vec{\Psi}}_{\vec{k}}^0
\end{equation}
for matrix $\eta\mathbb{H}_{\vec{k}}$. Note that although $\eta\mathbb{H}_{\vec{k}}$ is not a Hermitian matrix, $\eta\mathbb{H}_{\vec{k}}$ is a Hermitian operator in Hilbert space with pseudo-Euclidean metric $\eta$. Therefore, the eigenvalues $\varepsilon$ are real and the eigenvectors $\hat{\vec{\Psi}}^0=[\hat{\psi}^0_1(\vec{k}),\hat{\psi}^0_2(\vec{k}),\hat{\psi}^0_3(\vec{k}),\hat{\psi}^{0*}_1(-\vec{k}),\hat{\psi}^{0*}_2(-\vec{k}),\hat{\psi}^{0*}_3(-\vec{k})]^{T}$ form an orthogonal basis (for the nondegenerate case). The positive eigenvalues $\varepsilon_{\nu,\vec{k}}$ are presented in Eq.~\eqref{eq:E_k}.
 The remaining three eigenvalues of EVP~\eqref{eq:EVP} are negative. They correspond to nonphysical directions of precession of magnetic moments $\vec{m}_\ell$ around their equilibrium directions  $\vec{m}^0_\ell$. Therefore, they will not be considered here. In detail, dispersion relations \eqref{eq:E_k} are shown in Fig.~\ref{fig:disp} for two different values of magnetic field $B$.

Substituting \eqref{eq:FT} into \eqref{eq:HP}, in the linear approximation, we write
\begin{equation}
    \begin{split}&\vec{m}_\ell(\vec{R}^\ell_{\vec{n}})\approx\vec{m}^0_\ell\\
    &+\sqrt{\frac{2}{N}}\sum\limits_{\vec{k}\in\text{1.BZ}}\left[\vec{e}^+_\ell\hat{\psi}_\ell(\vec{k})+\vec{e}^-_\ell\hat{\psi}^*_\ell(-\vec{k})\right]e^{i\vec{k}\cdot\vec{R}_{\vec{n}}^\ell}.
    \end{split}
\end{equation}
If a spin wave with the wave vector $\vec{k}$ is excited, then the single-spin dynamics can be reconstructed as
\begin{align}\label{eq:reconstr}
    &\vec{m}_\ell(\vec{R}^\ell_{\vec{n}})\approx\vec{m}^0_\ell\\
    \nonumber&+\sqrt{\frac{2}{N}}\mathrm{Re}\left\{\left[\vec{e}^+_\ell\hat{\psi}_\ell^0(\vec{k})+\vec{e}^-_\ell\hat{\psi}_\ell^{0*}(-\vec{k})\right]e^{i(\vec{k}\cdot\vec{R}_{\vec{n}}^\ell-\omega t)}\right\},
\end{align}
where eigenfrequencies $\omega=\varepsilon/\hbar$ and components of the eigenvectors $\hat{\psi}_\ell^0(\vec{k})$ and $\hat{\psi}_\ell^{0*}(-\vec{k})$ are found from the EVP~\eqref{eq:EVP}.

\section{The absence of the in-plane component for the magnon magnetic moment}\label{app:mu-ip}

Let us first find the component of the magnon magnetic moment along the 1st sub-lattice. To this end, we apply the magnetic field $\vec{B}=B\vec{m}^0_1$, where $\vec{m}^0_1$ is determined by \eqref{eq:grnd} for the case $\Theta=\pi/2$. 
In this case, magnetic Hamiltonian $\mathcal{H}=\mathcal{H}_{\textsc{afm}}+\mathcal{H}_{\textsc{z}}$ has the same exchange contribution $\mathcal{H}_{\textsc{afm}}$ as in \eqref{eq:H-det}, but Zeeman part is different $\mathcal{H}_{\textsc{z}}=-B\mu_s\sum_{\vec{R}_{\vec{n}}}\bigl[\vec{m}_1(\vec{R}_{\vec{n}})+\vec{m}_2(\vec{R}_{\vec{n}}+\vec{\delta}_1)+\vec{m}_3(\vec{R}_{\vec{n}}+2\vec{\delta}_1)\bigr]\cdot\vec{m}^0_1$.
Assuming the coplanar ground state, we obtain the following energy per magnetic unit cell $\mathcal{H}_{\textsc{muc}}=3J[(1-3b)\cos(\Phi_1-\Phi_2)+\cos(\Phi_2-\Phi_3)+(1-3b)\cos(\Phi_3-\Phi_1)-3b]$,
where $b=B/B_0$. The equilibrium values of $\Phi_\nu$ are as follows
\begin{equation}\label{eq:Phi-nu}
    \Phi_1=\varphi_0,\quad\Phi_2=\frac{2\pi}{3}+\varphi_0-\varphi,\quad\Phi_3=\frac{4\pi}{3}+\varphi_0+\varphi,
\end{equation}
where $\varphi$ is determined by the equation
\begin{equation}\label{eq:var-phi}
    (1-3b)\cos\left(\frac{\pi}{6}-\varphi\right)=\cos\left(\frac{\pi}{6}+2\varphi\right).
\end{equation}
For $b\ll1$, Eq.~\eqref{eq:var-phi} has solution $\varphi\approx\sqrt{3}b$, thus $(\partial\varphi/\partial B)_{B=0}=\mu_s/(3\sqrt{3}J)$. 
The corresponding component of the magnon magnetic moment is 
\begin{equation}
 \nonumber\vec{\mu}_{\nu,\vec{k}}\cdot\vec{m}^0_1=-\left.\frac{\partial\varepsilon_{\nu,\vec{k}}}{\partial\varphi}\frac{\partial\varphi}{\partial B}\right|_{B=0}=-\frac{\mu_s}{2\sqrt{3}J}\left.\frac{\partial\varepsilon_{\nu,\vec{k}}}{\partial\varphi}\right|_{B=0}.   
\end{equation}
Note that $\vec{\mu}_{\nu,\vec{k}}\cdot\vec{m}^0_1=0$ if $\varepsilon_{\nu,\vec{k}}=\varepsilon_{\nu,\vec{k}}(\varphi^2)$, because $\varphi|_{B=0}=0$.

In order to find the magnon spectra, we utilize Holstein-Primakoff representation \eqref{eq:HP} with $\vec{m}^0_\nu=\vec{e}_x\cos\Phi_\nu+\vec{e}_y\sin\Phi_\nu$ and $\vec{e}^\pm_\nu=\frac12[\mp i \sin\Phi_\nu\vec{e}_x\pm i\cos\Phi_\nu\vec{e}_y-\vec{e}_z]/\sqrt{S}$, where angles $\Phi_\nu$ are determined in \eqref{eq:Phi-nu}. In terms of $\hat{\psi}_\nu(\vec{k})$, the harmonic part of the Hamiltonian is
    \begin{align}\label{eq:H2-inpl}
    \nonumber&\mathcal{H}^{(2)}=\varepsilon_0\sum\limits_{\vec{k}\in\text{1.BZ}}\Bigl\{\frac12\left[|\hat{\psi}_1(\vec{k})|^2+|\hat{\psi}_2(\vec{k})|^2+|\hat{\psi}_3(\vec{k})|^2\right]\\\nonumber
    &+\mathcal{A}_\varphi\alpha_{\vec{k}}\left[\hat{\psi}_1(\vec{k})\hat{\psi}^*_2(\vec{k})+\hat{\psi}_3(\vec{k})\hat{\psi}^*_1(\vec{k})\right]\\
    &+\mathcal{B}_\varphi\alpha_{\vec{k}}\left[\hat{\psi}_1(\vec{k})\hat{\psi}_2(-\vec{k})+\hat{\psi}_3(\vec{k})\hat{\psi}_1(-\vec{k})\right]\\\nonumber
    &+\mathcal{A}'_\varphi\alpha_{\vec{k}}\hat{\psi}_2(\vec{k})\hat{\psi}^*_3(\vec{k})+\mathcal{B}'_\varphi\alpha_{\vec{k}}\hat{\psi}_2(\vec{k})\hat{\psi}_3(-\vec{k})+\text{c.c.}\Bigr\},
\end{align}
where $\mathcal{A}_\varphi=\frac12[1-\sin(\frac{\pi}{6}-\varphi)]$, $\mathcal{A}'_\varphi=\frac12[1-\sin(\frac{\pi}{6}+2\varphi)]$, $\mathcal{B}_\varphi=\frac12[1+\sin(\frac{\pi}{6}-\varphi)]$, $\mathcal{B}'_\varphi=\frac12[1+\sin(\frac{\pi}{6}+2\varphi)]$. Writing \eqref{eq:H2-inpl}, we utilized the relation between $b$ and $\varphi$ determined by Eq.~\eqref{eq:var-phi}. The corresponding eigenvalue problem coincides with \eqref{eq:EVP}, but the form of matrices $\mathbb{A}_{\vec{k}}$ and $\mathbb{B}_{\vec{k}}$:
\begin{align}
    \nonumber&\mathbb{A}_{\vec{k}}=\begin{bmatrix}
        1 & \mathcal{A}_\varphi\alpha^*_{\vec{k}} & \mathcal{A}_\varphi\alpha_{\vec{k}} \\
        \mathcal{A}_\varphi\alpha_{\vec{k}} & 1 & \mathcal{A}'_\varphi\alpha^*_{\vec{k}} \\
        \mathcal{A}_\varphi\alpha^*_{\vec{k}} & \mathcal{A}'_\varphi\alpha_{\vec{k}} & 1
    \end{bmatrix},\\
    &\mathbb{B}_{\vec{k}}=\begin{bmatrix}
        0 & \mathcal{B}_\varphi\alpha^*_{\vec{k}} & \mathcal{B}_\varphi\alpha_{\vec{k}} \\
        \mathcal{B}_\varphi\alpha_{\vec{k}} & 0 & \mathcal{B}'_\varphi\alpha^*_{\vec{k}} \\
        \mathcal{B}_\varphi\alpha^*_{\vec{k}} & \mathcal{B}'_\varphi\alpha_{\vec{k}} & 0
    \end{bmatrix}
\end{align}
For $\varphi\ll1$, the corresponding positive eigenenergies are
\begin{subequations}\label{eq:disp-ipB}
    \begin{align}
    &\varepsilon_{\pm1,\vec{k}}=\varepsilon_0\left[\sqrt{1-\frac12\Omega^\pm_{\vec{k}}\left(\Omega^\pm_{\vec{k}}+1\right)}+\mathcal{O}(\varphi^2)\right],\\ 
& \varepsilon_{0,\vec{k}}=\varepsilon_0\left[\sqrt{\left(1-\Omega^c_{\vec{k}}\right)\left(1+2\Omega^c_{\vec{k}}\right)}+\mathcal{O}(\varphi^2)\right].
    \end{align}
\end{subequations}
The absence of the terms linear in $\varphi$ results in $\vec{\mu}_{\nu,\vec{k}}\cdot\vec{m}^0_1=0$, as it was discussed above. Since the first sublattice is not distinguished, we conclude that the projections of the magnon magnetic moment onto all three directions $\vec{m}^0_1$, $\vec{m}^0_2$, and $\vec{m}^0_3$ are also zero. From this, it follows that the in-plane component of $\vec{\mu}_{\nu,\vec{k}}$ vanishes.

\section{Spin-lattice simulations}\label{app:simuls}
The dynamics of magnetic moments are described by the Landau--Lifshitz equations

\begin{equation}\label{eq:sim_LL}
    \begin{split}
        \left(1+\alpha^2_\textsc{g}\right)\dot{\vec{m}}(\vec{r}_{\vec{n}})=\frac{\gamma}{\mu_s}\,&\vec{m}(\vec{r}_{\vec{n}})\times\frac{\partial\mathcal{H}}{\partial\vec{m}(\vec{r}_{\vec{n}})}\\
        +\alpha_\textsc{g}\,\vec{m}(\vec{r}_{\vec{n}})\times\biggl[&\vec{m}(\vec{r}_{\vec{n}})\times\frac{\partial\mathcal{H}}{\partial\vec{m}(\vec{r}_{\vec{n}})}\biggr],        
    \end{split}
\end{equation} 
where $\alpha_\textsc{g}$ is the Gilbert damping parameter and $\mathcal{H}$ is defined in~\eqref{eq:H}.
The dynamical problem is considered as a set of $3N_1 N_2$ ordinary differential equations~\eqref{eq:sim_LL} with respect to $3N_1 N_2$ unknown functions $m^\textsc{x}(t),\ m^\textsc{y}(t),\ m^\textsc{z}(t)$. Parameters $N_1$ and $N_2$ define the size of the system. For the given size and initial conditions, the set of time evolution equations~\eqref{eq:sim_LL} is integrated numerically using the Runge--Kutta method in Python with integration step $3\Delta t J\gamma/\mu_s \ll 1$.

To simulate the spin waves, we consider a system with a size of $N_1\times N_2 = 291\times336$ magnetic moments. The simulations are carried out in two steps. In the first step, we simulate the dynamics of the system in the external magnetic field $\vec{B}=B\vec{e}_z$ with $B/B_0\in\left[0,0.3\right]$. As an initial state, here we considered an ``umbrella'' state with sinc modulation, i.e. $\vec{m}_\ell = \frac{\vec{m}^0_\ell+\vec{\delta m}}{|\vec{m}^0_\ell+\vec{\delta m}|}$, where $\vec{m}^0_\ell$ is defined in \eqref{eq:grnd} and $\vec{\delta m} = 0.05\,\text{sinc}\left(2\pi\,\vec{k}\cdot\vec{r}_{\vec{n}}\right)(\vec{a}_\ell+\vec{b}_\ell)$.

In the second step, we perform a space-time Fourier transformation for the complex-valued parameter $m^\textsc{x}(t)+i m^\textsc{y}(t)$. The resulting eigenfrequencies are plotted in Fig.~\ref{fig:Zeeman}.

\section{Magnon spectra for a multilayer 3D generalization}\label{app:3D}

Here, we consider a 3D generalization of the proposed model in the form of a stack of infinitely many monolayers, which are considered above and shown in Fig.~\ref{fig:lattice}. The distance between layers is $c_0$. The layers interact antiferromagnetically with the exchange constant $J_z>0$. In magnetic field $\vec{B}=B\vec{e}_z$ the ground state is $\vec{m}^0_\ell=\sin\bar{\Theta}(\vec{e}_x\cos\Phi_\ell+\vec{e}_y\sin\Phi_\ell)+\cos\bar{\Theta}\vec{e}_z$ and $\vec{m}'^0_{\ell}=-\sin\bar{\Theta}(\vec{e}_x\cos\Phi_\ell+\vec{e}_y\sin\Phi_\ell)+\cos\bar{\Theta}\vec{e}_z$ for the odd and even layers, respectively. Here 
\begin{equation}
    \cos\bar{\Theta}=\frac{B\mu_s}{9J+4J_z}=\varsigma\cos\Theta,
\end{equation}
where $\varsigma=(1+\frac49\frac{J_z}{J})^{-1}$.
The primitive magnetic unit cell is a prism composed of three vectors $\vec{e}_1$, $\vec{e}_2$, and $\vec{e}_3=2c_0\vec{e}_z$. The number of sub-lattices is doubled and equal to six: $\vec{m}_1$, $\vec{m}_2$, $\vec{m}_3$, $\vec{m}'_{1}$, $\vec{m}'_{2}$, $\vec{m}'_{3}$. In the following, we perform the same calculation technique as described in Appendix~\ref{app:spectra}. The harmonic part of the 3D generalized Hamiltonian in $\vec{k}$-space is
\begin{widetext}
\begin{equation}\label{eq:H_3D_harmonic}
    \begin{split}\mathcal{H}^{(2)}_{3D}=&\varepsilon_0\sum\limits_{\vec{k}\in\text{1.BZ}}\Bigl\{\frac12\left[\hat{\psi}_\ell(\vec{k})\hat{\psi}^*_\ell(\vec{k})+\hat{\psi}'_{\ell}(\vec{k})\hat{\psi}'^*_{\ell}(\vec{k})\right]\\
    +\mathcal{A}\alpha_{\vec{k}}&\left[\hat{\psi}_1(\vec{k})\hat{\psi}^*_2(\vec{k})+\hat{\psi}_2(\vec{k})\hat{\psi}^*_3(\vec{k})+\hat{\psi}_3(\vec{k})\hat{\psi}^*_1(\vec{k})+\hat{\psi}'_{1}(\vec{k})\hat{\psi}'^*_{2}(\vec{k})+\hat{\psi}'_{2}(\vec{k})\hat{\psi}'^*_{3}(\vec{k})+\hat{\psi}'_{3}(\vec{k})\hat{\psi}'^*_{1}(\vec{k})\right]\\
    +\mathcal{B}\alpha_{\vec{k}}&\left[\hat{\psi}_1(\vec{k})\hat{\psi}_2(-\vec{k})+\hat{\psi}_2(\vec{k})\hat{\psi}_3(-\vec{k})+\hat{\psi}_3(\vec{k})\hat{\psi}_1(-\vec{k})+\hat{\psi}'_{1}(\vec{k})\hat{\psi}'_{2}(-\vec{k})+\hat{\psi}'_{2}(\vec{k})\hat{\psi}'_{3}(-\vec{k})+\hat{\psi}'_{3}(\vec{k})\hat{\psi}'_{1}(-\vec{k})\right]\\
    +j&\left[\frac12\cos2\bar\Theta\left[\hat{\psi}_\ell(\vec{k})\hat{\psi}^*_\ell(\vec{k})+\hat{\psi}'_{\ell}(\vec{k})\hat{\psi}'^*_{\ell}(\vec{k})\right]-\cos \left(k_z c_0\right)\hat{\psi}_\ell(\vec{k})\left[\sin^2\bar{\Theta}\hat{\psi}'_{\ell}(-\vec{k})-\cos^2\bar{\Theta}\hat{\psi}'^*_{\ell}(\vec{k})\right]\right] +\text{c.c.}\Bigr\},
    \end{split}
\end{equation}
where $j=2J_z/(3J)$, $\mathcal{A}$ and $\mathcal{B}$ are the same as in \eqref{eq:H2}, and summation over the repeating indices $\ell$ is assumed. The corresponding eigenvalue problem coincides with \eqref{eq:Schrod}--\eqref{eq:EVP}, but the form of matrices $\mathbb{A}_{\vec{k}}$ and $\mathbb{B}_{\vec{k}}$:
\begin{equation}
    \mathbb{A}_{\vec{k}}=\begin{bmatrix}
        \mathcal{F} & \mathcal{A}^*\alpha^*_{\vec{k}} & \mathcal{A}\alpha_{\vec{k}} & -\mathcal{C} & 0 & 0\\
        \mathcal{A}\alpha_{\vec{k}} & \mathcal{F} & \mathcal{A}^*\alpha^*_{\vec{k}} & 0 & -\mathcal{C} & 0\\
        \mathcal{A}^*\alpha^*_{\vec{k}} & \mathcal{A}\alpha_{\vec{k}} & \mathcal{F} & 0 & 0 & -\mathcal{C}\\
        -\mathcal{C} & 0 & 0 & \mathcal{F} & \mathcal{A}^*\alpha^*_{\vec{k}} & \mathcal{A}\alpha_{\vec{k}}\\
        0 & -\mathcal{C} & 0 & \mathcal{A}\alpha_{\vec{k}} & \mathcal{F} & \mathcal{A}^*\alpha^*_{\vec{k}}\\
        0 & 0 & -\mathcal{C} & \mathcal{A}^*\alpha^*_{\vec{k}} & \mathcal{A}\alpha_{\vec{k}} & \mathcal{F}
    \end{bmatrix},\qquad \mathbb{B}_{\vec{k}}=\begin{bmatrix}
        0 & \mathcal{B}\alpha^*_{\vec{k}} & \mathcal{B}\alpha_{\vec{k}} & \mathcal{S} & 0 & 0\\
        \mathcal{B}\alpha_{\vec{k}} & 0 & \mathcal{B}\alpha^*_{\vec{k}} & 0 & \mathcal{S} & 0\\
        \mathcal{B}\alpha^*_{\vec{k}} & \mathcal{B}\alpha_{\vec{k}} & 0 & 0 & 0 & \mathcal{S}\\
        \mathcal{S} & 0 & 0 & 0 & \mathcal{B}\alpha^*_{\vec{k}} & \mathcal{B}\alpha_{\vec{k}}\\
        0 & \mathcal{S} & 0 & \mathcal{B}\alpha_{\vec{k}} & 0 & \mathcal{B}\alpha^*_{\vec{k}}\\
        0 & 0 & \mathcal{S} & \mathcal{B}\alpha^*_{\vec{k}} & \mathcal{B}\alpha_{\vec{k}} & 0 
    \end{bmatrix},
\end{equation}
where $\mathcal{F} = 1+j$, $\mathcal{C} = j \cos\left(k_z c_0\right)\cos^2\bar{\Theta}$, and $\mathcal{S} = j\cos\left(k_z c_0\right)\sin^2\bar{\Theta}$. Thus, $\mathbb{H}_{\vec{k}}$ is $12\times12$ matrix, as well as metric $\eta=\mathrm{diag}(1,1,\dots,1,-1,-1,\dots,-1)$, which contains six `$1$' and another six `$-1$' along the diagonal. The corresponding six magnon branches have the following dispersion 
\begin{align}\label{eq:disp_j}
    \nonumber&\varepsilon_{\pm1,\vec{k}}=\varepsilon_0\left\{\sqrt{\left(1+\frac{\Omega^\pm_{\vec{k}}}{2}+2j\cos^2\frac{q_z}{2}\right)\left[1-\Omega^\pm_{\vec{k}}\left(1-\frac32\cos^2\bar{\Theta}\right)+j(1+\cos2\bar{\Theta}\cos q_z)\right]}\pm\cos\bar{\Theta}\left(\frac{\Omega^\pm_{\vec{k}}}{2}+\Omega^\mp_{\vec{k}}\right)\right\},\\
    \nonumber&\varepsilon'_{\pm1,\vec{k}}=\varepsilon_0\left\{\sqrt{\left(1+\frac{\Omega^\pm_{\vec{k}}}{2}+2j\sin^2\frac{q_z}{2}\right)\left[1-\Omega^\pm_{\vec{k}}\left(1-\frac32\cos^2\bar{\Theta}\right)+j(1-\cos2\bar{\Theta}\cos q_z)\right]}\pm\cos\bar{\Theta}\left(\frac{\Omega^\pm_{\vec{k}}}{2}+\Omega^\mp_{\vec{k}}\right)\right\},\\
    \nonumber&\varepsilon_{0,\vec{k}}=\varepsilon_0\left\{\sqrt{\left(1-\Omega^c_{\vec{k}}+2j\sin^2\frac{q_z}{2}\right)\left[1+\Omega^c_{\vec{k}}\left(2-3\cos^2\bar{\Theta}\right)+j(1-\cos2\bar{\Theta}\cos q_z)\right]}+\sqrt{3}\cos\bar{\Theta}\Omega^s_{\vec{k}}\right\},\\
    &\varepsilon'_{0,\vec{k}}=\varepsilon_0\left\{\sqrt{\left(1-\Omega^c_{\vec{k}}+2j\cos^2\frac{q_z}{2}\right)\left[1+\Omega^c_{\vec{k}}\left(2-3\cos^2\bar{\Theta}\right)+j(1+\cos2\bar{\Theta}\cos q_z)\right]}+\sqrt{3}\cos\bar{\Theta}\Omega^s_{\vec{k}}\right\},
\end{align}
where $q_z=k_zc_0$. For the case $j=0$, one obtains $\varepsilon'_{\nu,\vec{k}}=\varepsilon_{\nu,\vec{k}}$. From \eqref{eq:disp_j}, we obtain the magnetic moments for the 3D system $\bar{\mu}^z_{\nu,\vec{k}}=-\partial\varepsilon_{\nu,\vec{k}}/\partial B|_{B=0}$, $\bar{\mu}'^z_{\nu,\vec{k}}=-\partial\varepsilon'_{\nu,\vec{k}}/\partial B|_{B=0}$, and find that $\bar{\mu}'^z_{\nu,\vec{k}}=\bar{\mu}^z_{\nu,\vec{k}}=\varsigma \mu^z_{\nu,\vec{k}}$.
\end{widetext}

\section{Nonlinear spin splitter and Edelstein effects}\label{app:transport}

Here we use the standard sharp-quasiparticle Boltzmann treatment, assuming that the nonlinear contributions to the spin Hamiltonian are suppressed due to the large spin. I.e., we assume that $S=|g|\mu_s/\mu_{\textsc{b}}\gg1$. The spin current (flow of the magnetic moment) \eqref{eq:J-gen} in the spin-splitter effect, as well as the total magnetic moment \eqref{eq:M} in the Edelstein effect, are determined by the nonequilibrium part of the distribution function $\delta n_{\nu,\vec{k}}$. The latter can be found from the steady-state Boltzmann transport equation, which, in our case, in the relaxation-time approximation, is as follows
\begin{equation}\label{eq:Bltz}
    \vec{v}_{\nu,\vec{k}}\cdot\vec{\nabla}n_{\nu,\vec{k}}=-\frac{\delta n_{\nu,\vec{k}}}{\tau_{\mathrm{rlx}}}.
\end{equation}
Here $n_{\nu,\vec{k}}=n_{\nu,\vec{k}}^0+\delta n_{\nu,\vec{k}}$ is the total distribution function with 
\begin{equation}
    n^0_{\nu,\vec{k}}=\frac{1}{e^{\varepsilon_{\nu,\vec{k}}/(k_{\textsc{b}}T)}-1}
\end{equation}
being the equilibrium part. In the limit $\delta n_{\nu,\vec{k}}\ll n^0_{\nu,\vec{k}}$, the Boltzmann equation \eqref{eq:Bltz} can be solved for $\delta n_{\nu,\vec{k}}$ perturbatively. To this end, we present 
\begin{equation}\label{eq:n-expans}
    \delta n_{\nu,\vec{k}}=\delta n^{(1)}_{\nu,\vec{k}}+\delta n^{(2)}_{\nu,\vec{k}}+\delta n^{(3)}_{\nu,\vec{k}}\dots,
\end{equation}
where $\delta n^{(1)}_{\nu,\vec{k}}\gg\delta n^{(2)}_{\nu,\vec{k}}\gg\delta n^{(3)}_{\nu,\vec{k}}\gg\dots$. The latter condition is valid if $|\vec{\nabla}T|\ll1/(\ell_{\text{rlx}}\partial_Tn^0_{\nu,\vec{k}})$ with $\ell_{\text{rlx}}$ being the mean free path. For large temperatures, this condition can be simplified to $|\vec{\nabla}T|\ll\varepsilon_0/(\ell_{\text{rlx}}k_{\textsc{b}})\approx\hbar/(\tau_{\text{rlx}}a_0k_{\textsc{b}})\approx\Gamma_{\vec{k}}/(a_0k_{\textsc{b}})$. By substituting \eqref{eq:n-expans} into \eqref{eq:Bltz} and equating terms of the same order of smallness, we obtain the cascade of equations, whose solutions are
\begin{equation}\label{eq:delta-n}
    \begin{split}
    &\delta n^{(1)}_{\nu,\vec{k}}=-\tau_{\text{rlx}}(v_{\nu,\vec{k}})_\alpha\partial_\alpha n^0_{\nu,\vec{k}},\\
    &\delta n^{(2)}_{\nu,\vec{k}}=\tau^2_{\text{rlx}}(v_{\nu,\vec{k}})_\alpha(v_{\nu,\vec{k}})_\beta\partial^2_{\alpha\beta} n^0_{\nu,\vec{k}},\\
    &\delta n^{(3)}_{\nu,\vec{k}}=-\tau^3_{\text{rlx}}(v_{\nu,\vec{k}})_\alpha(v_{\nu,\vec{k}})_\beta(v_{\nu,\vec{k}})_\gamma\partial^3_{\alpha\beta\gamma} n^0_{\nu,\vec{k}}.
    \end{split}
\end{equation}
With \eqref{eq:delta-n}, the expansion coefficients in \eqref{eq:J} can be written as
\begin{equation}
    \begin{split}
        &G_{\alpha\beta}=-\frac{\tau_{\mathrm{rlx}}}{V}\sum\limits_{\nu,\vec{k}}\bar{\mu}^z_{\nu,\vec{k}}(v_{\nu,\vec{k}})_\alpha(v_{\nu,\vec{k}})_\beta \partial_Tn^0_{\nu,\vec{k}},\\
        &G^{(2)}_{\alpha\beta\gamma}=\frac{\tau^2_{\mathrm{rlx}}}{V}\sum\limits_{\nu,\vec{k}}\bar{\mu}^z_{\nu,\vec{k}}(v_{\nu,\vec{k}})_\alpha(v_{\nu,\vec{k}})_\beta (v_{\nu,\vec{k}})_\gamma\partial_T^2n^0_{\nu,\vec{k}},
    \end{split}
\end{equation}
where the summation over $\nu$ runs over all six branches. We find that $G_{\alpha\beta}=0$ and the nonzero components of $G^{(2)}_{\alpha\beta\gamma}$ are $-G_{xxx}^{(2)}=G_{xyy}^{(2)}=G_{yxy}^{(2)}=G_{yyx}^{(2)}\equiv G^{(2)}$. Performing the transition $\sum_{\vec{k}}(\dots)\to V\int\frac{\dd\vec{k}}{(2\pi)^3}(\dots)$, we write $G^{(2)}$ in form \eqref{eq:G2}, where the dimensionless $\mathcal{G}$ is as follows
\begin{equation}\label{eq:G-cal}
    \mathcal{G}(\mathcal{T},j_z)=-\varsigma\sum\limits_\nu\!\int\!\!\frac{\dd\vec{q}}{(2\pi)^3}\tilde{\mu}^z_{\nu,\vec{q}}(\tilde{v}_{\nu,\vec{q}})^3_x\frac{f_2(\tilde{\varepsilon}_{\nu,\vec{q}}/\mathcal{T})}{\tilde{\varepsilon}_{\nu,\vec{q}}^2}.
\end{equation}
Here $\mathcal{T}=k_{\textsc{b}}T/\varepsilon_0$ denotes the dimensionless temperature, and $j_z=J_z/J$. The integration is performed over the dimensionless wave vector with components $q_{x,y}=a_0k_{x,y}$ and $q_z=c_0k_z$. Here $q_z\in[-\pi/2,\pi/2]$ and the horizontal cross-section of the dimensionless Brillouin zone is shown in Fig.~\ref{fig:spct_3D}(c,d) by the gray shadowing. $\tilde{\mu}^z_{\nu,\vec{q}}=\bar{\mu}^z_{\nu,\vec{k}}/(|g|\mu_{\textsc{b}})$ and $\tilde{\varepsilon}_{\nu,\vec{q}}=\varepsilon_{\nu,\vec{k}}/\varepsilon_0$ are dimensionless magnon magnetic moment and energy, respectively. $\tilde{\vec{v}}_{\nu,\vec{q}}=\partial_{\vec{q}}\tilde{\varepsilon}_{\nu,\vec{q}}$ denotes the dimensionless group velocity. Fo the function $f_2$ in \eqref{eq:G-cal} we use the following definition
\begin{equation}\label{eq:f-def}
    f_n(x)=\hat{\mathbb{L}}_x^n\frac{1}{e^x-1},\qquad\hat{\mathbb{L}}_x=x^2\partial_x.
\end{equation}

The coefficients in expansion \eqref{eq:M-exp} are as follows
\begin{align}
    \nonumber &F_{\alpha}=-\frac{\tau_{\mathrm{rlx}}}{V}\sum\limits_{\nu,\vec{k}}\bar{\mu}^z_{\nu,\vec{k}}(v_{\nu,\vec{k}})_\alpha \partial_Tn^0_{\nu,\vec{k}},\\
        &F^{(2)}_{\alpha\beta}=\frac{\tau^2_{\mathrm{rlx}}}{V}\sum\limits_{\nu,\vec{k}}\bar{\mu}^z_{\nu,\vec{k}}(v_{\nu,\vec{k}})_\alpha(v_{\nu,\vec{k}})_\beta \partial^2_Tn^0_{\nu,\vec{k}},\\
        \nonumber&F^{(3)}_{\alpha\beta\gamma}=-\frac{\tau^3_{\mathrm{rlx}}}{V}\sum\limits_{\nu,\vec{k}}\bar{\mu}^z_{\nu,\vec{k}}(v_{\nu,\vec{k}})_\alpha(v_{\nu,\vec{k}})_\beta (v_{\nu,\vec{k}})_\gamma\partial_T^3n^0_{\nu,\vec{k}}.
\end{align}
We find that $F_\alpha=0$, $F^{(2)}_{\alpha\beta}=0$ and nonzero components of tensor $F^{(3)}_{\alpha\beta\gamma}$ are related as $F_{xxx}^{(3)}=-F_{xyy}^{(3)}=-F_{yxy}^{(3)}=-F_{yyx}^{(3)}\equiv F^{(3)}$, where $F^{(3)}$ is defined in \eqref{eq:F3} with
\begin{equation}\label{eq:F-cal}
    \mathcal{F}(\mathcal{T},j_z)=\varsigma\sum\limits_\nu\!\int\!\!\frac{\dd\vec{q}}{(2\pi)^3}\tilde{\mu}^z_{\nu,\vec{q}}(\tilde{v}_{\nu,\vec{q}})^3_x\frac{f_3(\tilde{\varepsilon}_{\nu,\vec{q}}/\mathcal{T})}{\tilde{\varepsilon}_{\nu,\vec{q}}^3}.
\end{equation}

From definition \eqref{eq:f-def}, it follows that $\partial_xf_n(x)=f_{n+1}(x)/x^2$. Using this, we derive
\begin{equation}
    \partial_{\mathcal{T}}\frac{f_n(\tilde{\varepsilon}_{\nu,\vec{q}}/\mathcal{T})}{\tilde{\varepsilon}_{\nu,\vec{q}}^n}=-\frac{f_{n+1}(\tilde{\varepsilon}_{\nu,\vec{q}}/\mathcal{T})}{\tilde{\varepsilon}_{\nu,\vec{q}}^{n+1}}.
\end{equation}
Now, comparing \eqref{eq:G-cal} and \eqref{eq:F-cal} we conclude that $\mathcal{F}=\partial_{\mathcal{T}}\mathcal{G}$.

\end{document}